%% file: main.tex
\documentclass[10pt,conference]{IEEEtran}
\usepackage{cite}
\usepackage{amsmath,amssymb,amsfonts}
\usepackage{algorithmic}
\usepackage{graphicx}
\usepackage{textcomp}
\usepackage{xcolor}
\usepackage[hyphens]{url}
\usepackage{fancyhdr}
\usepackage{hyperref}
\input{imports}

\newcommand{\hpcayear}{2025}

\newcommand{\hpcasubmissionnumber}{1586}
\title{Uni-Render: A Unified Accelerator for Real-Time Rendering Across Diverse Neural Renderers} 

\def\hpcacameraready{} 

\newcommand\hpcaauthors{Chaojian Li, Sixu Li, Linrui Jiang, Jingqun Zhang, and Yingyan (Celine) Lin}
\newcommand\hpcaaffiliation{Georgia Institute of Technology}
\newcommand\hpcaemail{\{cli851, sli941, jzhang3368, celine.lin\}@gatech.edu}

\input{hpca-template}

\begin{document}
\maketitle

\ifdefined\hpcacameraready 
  \thispagestyle{camerareadyfirstpage}
  \pagestyle{empty}
\else
  \thispagestyle{plain}
  \pagestyle{plain}
\fi

\newcommand{\hpcaheight}{0mm}
\ifdefined\eaopen
\renewcommand{\hpcaheight}{12mm}
\fi

\input{sections/0-Abstract}
\input{sections/1-Introduction}
\input{sections/2-Background}

\input{sections/3-Benchmark}

\input{sections/4-Method}

\input{sections/5-Experiments}
\input{sections/6-RelatedWorks}
\input{sections/7-Conclusion}
\input{sections/8-Acknowledgment}

\bibliographystyle{IEEEtranS}
\bibliography{refs}

\end{document}

%% file: imports.tex
\usepackage{amsmath,amsfonts}
\usepackage{algorithmic}
\usepackage{graphicx}
\usepackage{textcomp}
\usepackage{xcolor}
\usepackage{url}
\usepackage{hyperref}
\usepackage{wrapfig,lipsum,booktabs}
\usepackage{mathtools}
\usepackage{multirow}
\usepackage{makecell}

\usepackage{epsfig}
\usepackage{fancyhdr}
\usepackage[normalem]{ulem}
\usepackage{soul}

\usepackage{pifont}
\usepackage[textsize=footnotesize,textwidth=1.05\linewidth]{todonotes}

\makeatletter
\if@todonotes@disabled

\else

\fi
\makeatother

\usepackage{MnSymbol}

\usepackage[caption=false]{subfig}

\usepackage{stfloats}
\usepackage{xcolor,pifont,colortbl}
\definecolor{mygreen}{HTML}{009800}
\definecolor{darkgreen}{HTML}{009800}
\definecolor{lightgreen}{HTML}{B9E0A5}
\definecolor{lightyellow}{HTML}{F8CECC}
\definecolor{lightred}{HTML}{FF0000}
\newcommand*\colourcheck[1]{%
  \expandafter\newcommand\csname #1check\endcsname{\textcolor{#1}{\ding{52}}}%
}
\newcommand*\colourcross[1]{%
  \expandafter\newcommand\csname #1cross\endcsname{\textcolor{#1}{\ding{56}}}%
}
\colourcheck{mygreen}
\colourcross{red}

%% file: hpca-template.tex
\author{
  \ifdefined\hpcacameraready
    \IEEEauthorblockN{\hpcaauthors{}}
      \IEEEauthorblockA{
        \hpcaaffiliation{} \\
        \hpcaemail{}
      }
  \else
    \IEEEauthorblockN{\normalsize{HPCA \hpcayear{} Submission
      \textbf{\#\hpcasubmissionnumber{}}} \\
      \IEEEauthorblockA{
        Confidential Draft \\
        Do NOT Distribute!!
      }
    }
  \fi 
}

\fancypagestyle{camerareadyfirstpage}{%
  \fancyhead{}
  
  \fancyhead[C]{
    \ifdefined\aeopen
    \parbox[][12mm][t]{13.5cm}{\hpcayear{} IEEE International Symposium on High-Performance Computer Architecture (HPCA)}    
    \else
      \ifdefined\aereviewed
      \parbox[][12mm][t]{13.5cm}{\hpcayear{} IEEE International Symposium on High-Performance Computer Architecture (HPCA)}
      \else
      \ifdefined\aereproduced
      \parbox[][12mm][t]{13.5cm}{\hpcayear{} IEEE International Symposium on High-Performance Computer Architecture (HPCA)}
      \else
      \parbox[][0mm][t]{13.5cm}{\hpcayear{} IEEE International Symposium on High-Performance Computer Architecture (HPCA)}
    \fi 
    \fi 
    \fi 
    \ifdefined\aeopen 
      \includegraphics[width=12mm,height=12mm]{ae-badges/open-research-objects.pdf}
    \fi 
    \ifdefined\aereviewed
      \includegraphics[width=12mm,height=12mm]{ae-badges/research-objects-reviewed.pdf}
    \fi 
    \ifdefined\aereproduced
      \includegraphics[width=12mm,height=12mm]{ae-badges/results-reproduced.pdf}
    \fi
  }
  \fancyfoot[C]{}
}


%% file: sections/0-Abstract.tex
\begin{abstract}
Recent advancements in neural rendering technologies and their supporting devices have paved the way for immersive 3D experiences, significantly transforming human interaction with intelligent devices across diverse applications. However, achieving the desired real-time rendering speeds for immersive interactions is still hindered by (1) the lack of a universal algorithmic solution for different application scenarios and (2) the dedication of existing devices or accelerators to merely specific rendering pipelines. To overcome this challenge, we have developed a unified neural rendering accelerator that caters to a wide array of typical neural rendering pipelines, enabling real-time and on-device rendering across different applications while maintaining both efficiency and compatibility. Our accelerator design is based on the insight that, although neural rendering pipelines vary and their algorithm designs are continually evolving, they typically share common operators, predominantly executing similar workloads. Building on this insight, we propose a reconfigurable hardware architecture that can dynamically adjust dataflow to align with specific rendering metric requirements for diverse applications, effectively supporting both typical and the latest hybrid rendering pipelines. Benchmarking experiments and ablation studies on both synthetic and real-world scenes demonstrate the effectiveness of the proposed accelerator. It achieves real-time rendering speeds ($>$ 30 FPS) and up to 119$\times$ speedups over state-of-the-art neural rendering hardware across varied rendering pipelines, while adhering to power consumption constraints of around 5 W, typical for edge devices. Consequently, the proposed unified accelerator stands out as the first solution capable of achieving real-time neural rendering across varied representative pipelines on edge devices, potentially paving the way for the next generation of neural graphics applications.
\end{abstract}

%% file: sections/1-Introduction.tex
\section{Introduction}
\begin{figure}[!t]
     \centering
     \subfloat[Neural rendering~\cite{tewari2022advances,tewari2020state} incorporates \textbf{neural networks} into rendering pipelines to learn the physical parameters through gradient descents, alleviating the necessity of extensive manual inputs by 3D artists and enabling the generation of \textbf{photo-realistic} images.]{\includegraphics[width=\linewidth]{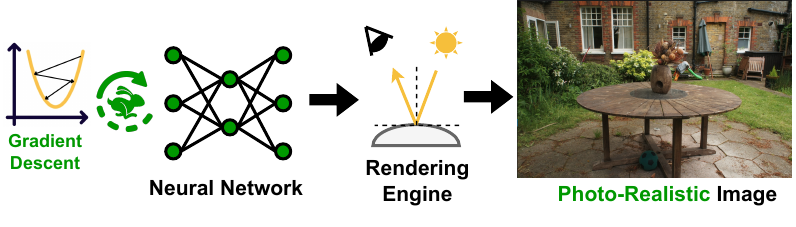}}
     \vspace{-0.0em}
     \subfloat[Examples of typical application scenarios of neural rendering~\cite{meka2019deep,philip2019multi,sun2019single,zhou2019deep,ma2021pixel,grassal2022neural,zielonka2023instant,lombardi2021mixture,ma2019accurate,liu2023mv,song2019apollocar3d,quest2,hesai}. 
     ]{\includegraphics[width=\linewidth]{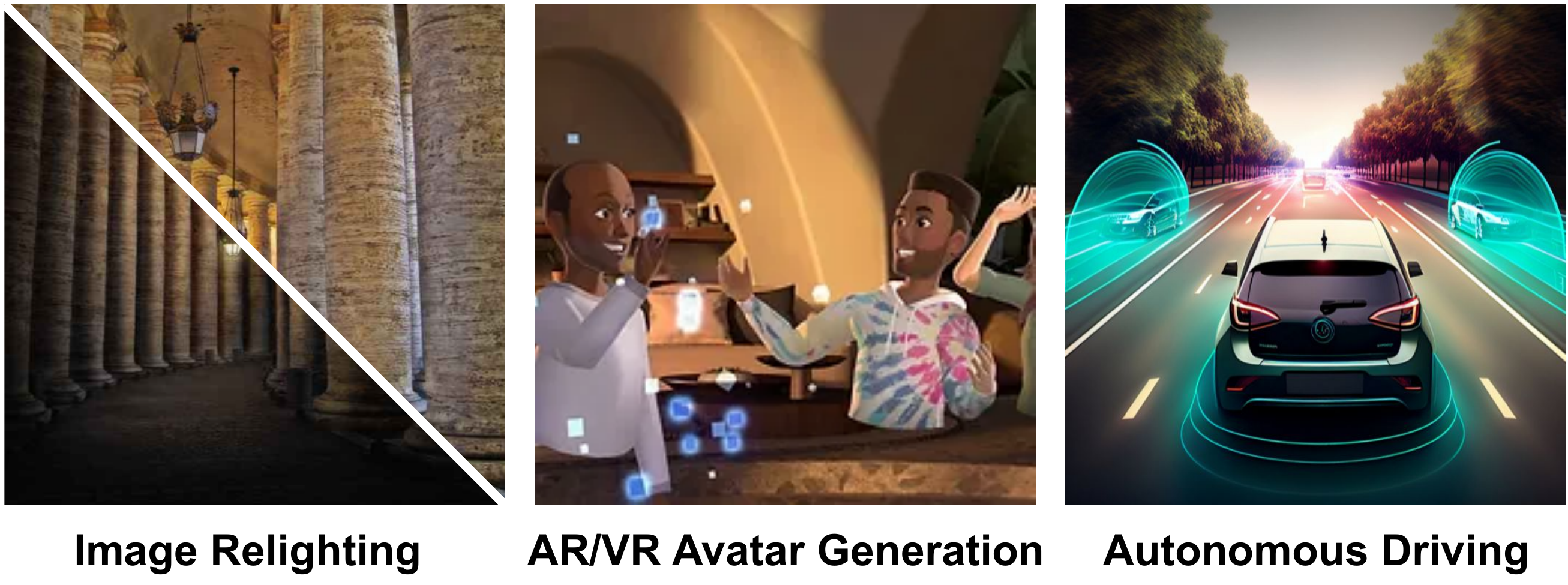}}
    \caption{
Illustrating the impact of neural rendering: (a) demonstrates how neural rendering facilitates the creation of photo-realistic images with reduced manual input compared to traditional rendering; and (b) highlights examples of various applications that can significantly benefit from neural rendering.
    }
    \vspace{-1.5em}
    \label{fig:traditional_vs_neural_rendering}
\end{figure}

The rapid advancements in neural rendering~\cite{tewari2022advances,tewari2020state}, alongside their supporting devices~\cite{questpro,vision_pro},
have promised an exciting era where 3D realistic experiences become the key interface between humans and intelligent devices, particularly in applications like social metaverse~\cite{meta_telepresence} and industrial design workflows~\cite{sharpr3d}, as depicted in Fig.~\ref{fig:traditional_vs_neural_rendering}. Despite the excitement and transformative potential of these developments, a notable challenge in fully realizing this potential is accommodating applications with diverse workloads~\cite{kwon2023xrbench}. Each often requires a unique set of rendering metrics~\cite{tewari2022advances,tewari2020state}, such as rendering quality, efficiency, storage size, and compatibility, while still maintaining real-time speeds of over 30 frames per second (FPS). This frame rate is crucial for immersive user interactions across a variety of emerging graphics applications, including novel view synthesis~\cite{hedman2018deep,sitzmann2019deepvoxels,aliev2020neural,mildenhall2020nerf}, image relighting~\cite{meka2019deep,philip2019multi,sun2019single,zhou2019deep}, AR/VR avatar generation~\cite{ma2021pixel,grassal2022neural,zielonka2023instant,lombardi2021mixture}, autonomous driving~\cite{ma2019accurate,liu2023mv,song2019apollocar3d}, and free-viewpoint video generation~\cite{lombardi2019neural,martin2018lookingood,pandey2019volumetric}.

The challenge of achieving desired real-time rendering speeds across diverse applications stems from two major barriers:
(1) Despite active algorithmic research efforts, it is unlikely that a single neural rendering pipeline capable of excelling across all crucial metrics required by diverse applications will emerge in the near future. As detailed in Sec.~\ref{sec:preliminaries}, no existing neural rendering pipeline currently surpasses others in every key metric simultaneously.
(2) No current neural rendering accelerator or device can effectively accelerate all representative neural rendering pipelines. This limitation prevents meeting the diverse requirements of different rendering metrics across various applications on individual devices. For example, as shown in Sec.~\ref{sec:benchmark}, Qualcomm 8Gen2 mobile development kits~\cite{qualcomm-8g2} achieve a 2.4$\times$ speedup over the Xavier NX edge GPU~\cite{xnx} for mesh-based pipelines~\cite{mobilenerf}, but are even 1.75$\times$ slower for low-rank-decomposed-grid-based pipelines~\cite{reiser2023merf}.
Additionally, dedicated accelerators~\cite{han2023metavrain,li2022rt,li2023instant} currently ensure real-time rendering speeds only for their targeted neural rendering pipelines.

The aforementioned barriers also resonate with the computer graphics (CG) community’s recent call for the development of neural rendering hardware accelerators, as emphasized: \textit{``modern hardware accelerators are designed for classical computer graphics and could in the future be similarly tailored to neural representations”}\cite{tewari2022advances}. Therefore, the motivation for designing a \textbf{unified neural rendering accelerator} can be justified from the following two perspectives: (1) On the \textbf{necessity}: In the rapidly evolving neural graphics field, it is even more crucial for the hardware architecture community to engage and contribute early. This enables offering hardware insights to inspire and motivate the algorithm community towards hardware-aware rendering pipelines.
(2) On the \textbf{feasibility}: An increasing number of rendering pipelines combine existing neural components\cite{li2024mixrt,lee2023compact,adaptiveshells2023,franke2024trips,morgenstern2023compact,chen2023dictionary}. This trend not only opens up opportunities but also strengthens the motivation to support diverse pipelines by accelerating the underlying common micro-operators.

To echo the aforementioned need for a unified neural rendering accelerator, we propose an accelerator called \textit{Uni-Render}, which is dedicated to the common micro-operators shared by varied neural rendering pipelines. This approach enables real-time on-device neural rendering, facilitating diverse rendering applications and workloads. Our contributions can be summarized as follows:

\begin{itemize}
    \item To identify the gap between current neural rendering hardware capabilities and the required real-time rendering speeds for diverse applications, we conducted extensive benchmarks and analyses on various edge devices and accelerators. Our key abstracted observations include: (1) Although supporting varied rendering pipelines is important for enabling diverse applications, not all rendering pipelines can achieve the desired real-time speeds on existing neural rendering hardware; (2) This inefficiency is due to the dedication of current devices and accelerators to specific pipelines rather than a broad spectrum of typical pipelines.
 
    \item Motivated by the aforementioned gap, we identified that both typical and the latest hybrid rendering pipelines \textit{can be decomposed into} a set of common micro-operators shared by varied rendering pipelines. Specifically, these common micro-operators execute similar workloads involving \textbf{indexing} tensor elements followed by \textbf{reduction} operations. The insight of this consistency presents an opportunity to develop a unified accelerator for varying rendering pipelines.
    
    \item Building upon our findings, we propose a reconfigurable hardware architecture that can adapt to varied neural rendering pipelines by dynamically adjusting the dataflow of different micro-operators to meet specific rendering metric requirements. Furthermore, such an architecture, designed for common micro-operators rather than specific rendering pipelines, ensures compatibility with the latest hybrid neural rendering pipelines.
    
    \item Benchmarking experiments and ablation studies on both synthetic~\cite{mildenhall2020nerf} and real-world~\cite{barron2022mip} scenes validate the efficiency of our proposed unified accelerator, demonstrating up to 119$\times$ speedups over state-of-the-art neural rendering hardware, across five typical rendering pipelines. To the best of our knowledge, this is \textbf{the first} accelerator to enable real-time ($>$ 30 FPS) on-device neural rendering across varied rendering pipelines while maintaining power consumption \textcolor{black}{around} 5 W, typical for edge devices~\cite{qualcomm-8g2, raspi4}.
\end{itemize}

%% file: sections/2-Background.tex
\begin{table*}[!t]
\caption{A comparative overview of typical rendering pipelines. The \textbf{rendering speed} is measured in FPS for executing the corresponding implementation with state-of-the-art accuracy vs. efficiency trade-offs~\cite{mobilenerf,instant-ngp,kilonerf,kerbl20233d} on the challenging Unbounded-360 dataset~\cite{barron2022mip}, using an edge GPU Orin NX~\cite{onx}, with a rendering resolution of 1280$\times$720~\cite{reiser2023merf,li2024mixrt}. The \textbf{rendering quality} and \textbf{storage efficiency} are cited from the corresponding works, which are measured by Peak Signal-to-Noise Ratio (PSNR) and required storage in megabytes (MB), respectively, across seven publicly accessible scenes from the Unbounded-360 dataset~\cite{barron2022mip}. The \textbf{CG toolchain compatibility} is assessed based on integration support in prevalent CG tools~\cite{haas2014history,blender,unrealengine,maya}.}
\vspace{-0.5em}
\centering
\setlength{\tabcolsep}{2pt}
  \resizebox{1.0\linewidth}{!}
  {
    \begin{tabular}{c|c||c|c|c|c||c}
    \toprule
    \textbf{Dominant Scene} & \textbf{Rendering} & \textbf{Rendering} & \textbf{Rendering} & \textbf{Storage} & \textbf{CG Toolchain} & \textbf{Representative} \\
    \textbf{Representation} & \textbf{Technique} & \textbf{Speed} & \textbf{Quality} & \textbf{Efficiency} & \textbf{Compatibility} & \textbf{Works} \\
    \midrule
    \multirow{2}{*}{\textbf{Mesh}} & \multirow{2}{*}{Rasterization} & \cellcolor{darkgreen}\textcolor{white}{Very High} & \cellcolor{lightred}\textcolor{white}{Low} & \cellcolor{lightyellow}Medium &  \cellcolor{darkgreen}\textcolor{white}{Very High} &  \multirow{2}{*}{\cite{mobilenerf,yariv2023bakedsdf,tang2023delicate,lin2023magic3d}} \\
     & & ($\leq$ 20 FPS on~\cite{onx}) & ($\leq$ 28 PSNR) & ($\leq$ 700 MB) & (\mygreencheck Unity~\mygreencheck Blender~\mygreencheck UE~\mygreencheck Maya) \\
    \midrule
    \multirow{2}{*}{\textbf{MLP}} & Volume  & \cellcolor{lightred}\textcolor{white}{Low} & \cellcolor{darkgreen}\textcolor{white}{Very High} & \cellcolor{darkgreen}\textcolor{white}{Very High} & \cellcolor{lightred}\textcolor{white}{Low} & \multirow{2}{*}{\cite{mildenhall2020nerf,verbin2022ref,tancik2022block,turki2022mega}}\\
     & Rendering & ($\leq$ 0.2 FPS on~\cite{onx}) & ($\leq$ 33 PSNR) & ($\leq$ 40 MB) & (\mygreencheck Unity~\redcross Blender~\redcross UE~\redcross Maya) \\
     \midrule
    \textbf{Low-Rank} & Volume & \cellcolor{lightgreen}High & \cellcolor{lightyellow}Medium &  \cellcolor{lightgreen}High & \cellcolor{lightred}\textcolor{white}{Low} & \multirow{2}{*}{\cite{chen2022tensorf,reiser2023merf,chen2023factor,wang2023rodin}}\\
    \textbf{Decomposed Grid} & Rendering & ($\leq$ 10 FPS on~\cite{onx}) & ($\leq$ 29 PSNR) & ($\leq$ 160 MB) & (\mygreencheck Unity~\redcross Blender~\redcross UE~\redcross Maya) \\
    \midrule
    \multirow{2}{*}{\textbf{Hash Grid}} & Volume  & \cellcolor{lightyellow}Medium & \cellcolor{lightyellow}Medium & \cellcolor{lightgreen}High & \cellcolor{lightgreen}High & \multirow{2}{*}{\cite{instant-ngp,barron2023zip,klenk2023nerf,park2022instantxr}}\\
     & Rendering & ($\leq$ 1 FPS on~\cite{onx}) & ($\leq$ 30 PSNR) & ($\leq$ 110 MB) & (\mygreencheck Unity~\mygreencheck Blender~\mygreencheck UE~\redcross Maya) \\
    \midrule
    \multirow{2}{*}{\textbf{3D Gaussian}} & Splat-Based & \cellcolor{lightgreen}High & \cellcolor{darkgreen}\textcolor{white}{Very High} & \cellcolor{lightyellow}Medium & \cellcolor{lightgreen}High & \multirow{2}{*}{\cite{kerbl20233d,luiten2023dynamic,yang2023deformable,keselman2023flexible}}\\
     & Rasterization & ($\leq$ 5 FPS on~\cite{onx}) & ($\leq$ 32 PSNR) & ($\leq$ 600 MB) & (\mygreencheck Unity~\mygreencheck Blender~\mygreencheck UE~\redcross Maya) \\
    \bottomrule
    \end{tabular}
    }
  \label{tab:rendering_pipelines}
\vspace{-0.7em}
\end{table*}

\section{Comprehensive Analysis of Typical Neural Rendering Pipelines}
\label{sec:preliminaries}

As demonstrated in Fig.~\ref{fig:traditional_vs_neural_rendering} (a), neural rendering integrates neural networks into 3D rendering pipelines to enhance rendering quality. The two key factors distinguishing different neural rendering pipelines are: (1) the 3D \textbf{scene representations}, e.g., the ``neural networks'' in Fig.~\ref{fig:traditional_vs_neural_rendering} (a), which store 3D scene information; and (2) the \textbf{rendering techniques}, e.g., the ``rendering engine'' in Fig.~\ref{fig:traditional_vs_neural_rendering} (a), used to form images. The input to a ``rendering engine'' is the camera pose corresponding to the view that the user wants to observe the scene from, and the output is the 2D image corresponding to this camera pose.

Based on these distinguishing factors, we categorize commonly used neural rendering works into \textbf{five typical rendering pipelines}, each with varied scene representations and rendering techniques. As summarized in Tab.~\ref{tab:rendering_pipelines}, these pipelines are widely recognized as state-of-the-art\cite{tewari2022advances,ramamoorthi2023nerfs} in terms of the following four important metrics: (1) \textbf{Rendering speed}, crucial for providing immersive and smooth user experiences~\cite{beigbeder2004effects,sumbul2022system}; (2) \textbf{Rendering quality}, key to enabling the integration of virtual and real-world scenes~\cite{kronander2015photorealistic,agusanto2003photorealistic}; (3) \textbf{Storage efficiency}, a primary consideration for sharing neural rendering models in contexts with limited or unreliable internet access~\cite{turki2022mega,li2023instant}; and (4) \textbf{CG toolchain compatibility}, measuring the ease with which artists and researchers can incorporate the rendering models into existing commercial CG toolchains, such as Blender~\cite{blender} and Maya~\cite{maya}.

Note that while the comparison results in Tab.~\ref{tab:rendering_pipelines} may change over time, the applicability of our proposed unified accelerator will not be compromised. This is because the accelerator is designed to focus on accelerating micro-operators, into which the latest hybrid rendering pipelines can be decomposed. The detailed illustration of these micro-operators and the results on the latest hybrid pipelines are presented in Sec.~\ref{sec:ops} and Sec.~\ref{sec:exp_hybrid}, respectively.

\subsection{Mesh-Based Rendering Pipelines}
\label{sec:preliminaries:mesh}
\begin{figure}[b]
  \centering
  \vspace{-1.5em}
  \includegraphics[width=1\linewidth]{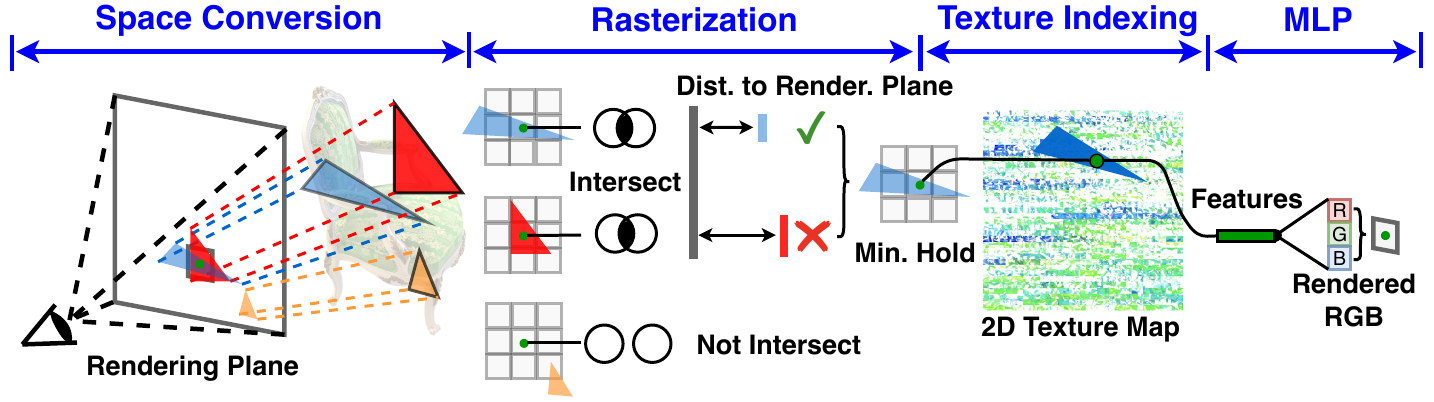}
\vspace{-1.5em}
\caption{An overview of \textbf{mesh-based} rendering pipelines. ``Dist. to Render. Plane” represents the distance between the ray-mesh intersection point and the rendering plane, while ``Min. Hold” refers to maintaining the minimal distance mentioned above and the corresponding mesh. Details on each step within these pipelines are provided in Sec.~\ref{sec:preliminaries:mesh}.}
\label{fig:mesh_pipeline}
\end{figure}

\textbf{Pipeline Overview.} 
Fig.~\ref{fig:mesh_pipeline} shows: mesh-based rendering pipelines consist of the following steps: (1) \underline{Space Conversion}: Polygonal meshes in world space are projected into clip space \cite{clip_space}, as determined by the view camera's rendering plane; (2) \underline{Rasterization}: This step traverses the projected meshes to identify the mesh closest to the rendering plane that intersects with the target pixel (e.g., the green dot in Fig.~\ref{fig:mesh_pipeline}) for rendering; (3) \underline{Texture Indexing}: The coordinates of the target pixel, mapped onto the identified mesh in the rasterization step (e.g., the blue triangle in Fig.~\ref{fig:mesh_pipeline}), are used to retrieve features from a 2D pre-computed or learned texture map through bilinear or nearest neighbor interpolation; (4) Multi-Layer Perceptron (\underline{MLP}): The retrieved features are input into a learned MLP, outputting the final rendered RGB color of the target pixel.

\textbf{Dominant Scene Representation.}
Mesh-based neural rendering pipelines predominantly utilize polygonal meshes (e.g., triangles) as their scene representations: These meshes serve as a piece-wise linear approximation of scene surfaces, storing critical information including (1) the coordinates of all vertices and (2) the indices of vertices forming each mesh. Rather than storing textures as vertex attributes~\cite{yariv2023bakedsdf}, it is more common to store them as 2D texture maps with a predefined mapping between the map and vertex locations \cite{mobilenerf,rakotosaona2024nerfmeshing}.

\textbf{Rendering Technique.} Rasterization~\cite{laine2011high} is the fundamental rendering technique in mesh-based rendering pipelines. After projecting all vertices into clip space using the space conversion step (shown in Fig.~\ref{fig:mesh_pipeline}), the information stored for each mesh is utilized to determine if the target pixel intersects with any mesh and to interpolate the corresponding depth. Subsequently, the mesh with the minimal depth at each pixel is stored in the buffer to store the depth value of each pixel, as outlined in ~\cite{laine2011high}.

\subsection{MLP-Based Rendering Pipelines}
\label{sec:preliminaries:mlp}

\begin{figure}[b]
\vspace{-1.5em}
  \centering
  \includegraphics[width=1\linewidth]{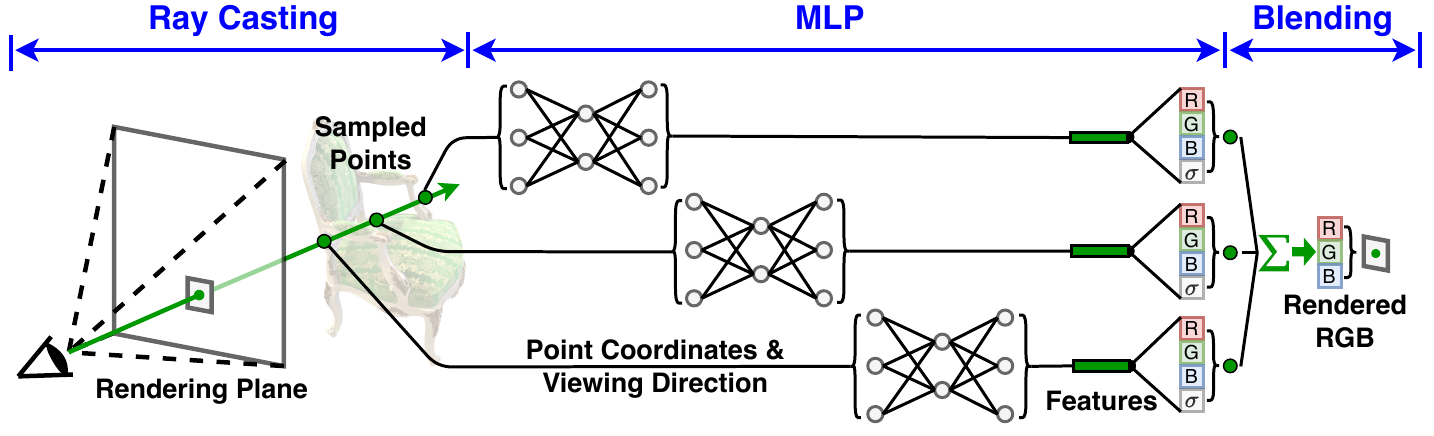}
\vspace{-2em}
\caption{An overview of \textbf{MLP-based} rendering. The detailed description of each step inside the pipelines is described in Sec.~\ref{sec:preliminaries:mlp}.}
\label{fig:mlp_pipeline}
\end{figure}

\textbf{Pipeline Overview.} As visualized in Fig.~\ref{fig:mlp_pipeline}, MLP-based rendering pipelines can be broken down to the following steps: (1) \underline{Ray Casting:} for the target pixel (e.g., the green dot in Fig.~\ref{fig:mlp_pipeline}), a ray is emitted/cast from the view camera and 3D points are sampled along the ray; (2) \underline{MLP}: the coordinates and view directions of the sampled points are sent as inputs to a learned MLP model to output the RGB color and density $\sigma$; (3) \underline{Blending}: following the principles of volume rendering~\cite{kajiya1984ray}, the colors and density of all sampled points along the ray corresponding the target pixel are blended to output the finally rendered RGB color.

\textbf{Dominant Scene Representation.} In MLP-based rendering pipelines, all the 3D scene information is retrieved from MLP in an implicit way (i.e., inputting the location of a arbitrary 3D point to the MLP and getting corresponding color and transparency information of this point). It is worth noting that although the MLP weights are the dominant scene representations, they cannot be explicitly visualized to show the 3D scene. However, as MLPs are known to act as universal function approximators~\cite{hornik1989multilayer}, MLP-based rendering pipelines can even enable the state-of-the-art rendering quality in city-scale scenes with the recent advances in deep learning~\cite{citynerf,tancik2022block}.

\textbf{Rendering Technique.} The key enabler in MLP-based rendering pipelines is the volume rendering technique~\cite{pandey2019volumetric}. 
In traditional graphics pipelines, this technique is mainly used for rendering participating media like fog, rather than rigid surfaces~\cite{pandey2019volumetric}. However, this technique does not set any assumptions about the objects' transparency; thus, it is widely used in MLP rendering pipelines for the ease of differentiation.

\subsection{Low-Rank-Decomposed-Grid-Based Rendering Pipelines}
\label{sec:preliminaries:low_rank}
\begin{figure}[b]
  \centering
   \vspace{-1.5em}
  \includegraphics[width=1\linewidth]{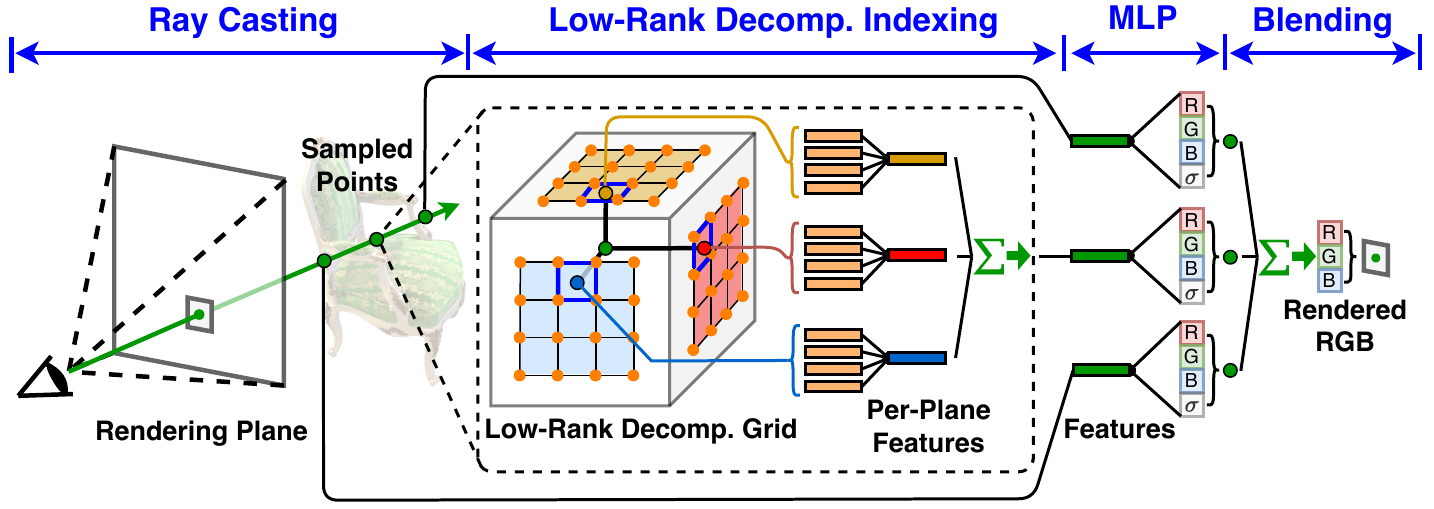}
\vspace{-2.2em}
\caption{An overview of \textbf{low-rank-decomposed-grid-based} rendering pipelines. The illustration of each step inside the pipelines is in Sec.~\ref{sec:preliminaries:low_rank}.}
\label{fig:low_rank_decomp_grid_pipeline}
\end{figure}

\textbf{Pipeline Overview.} As depicted in Fig.~\ref{fig:low_rank_decomp_grid_pipeline}, low-rank-decomposed-grid-based rendering pipelines are similar to the MLP-based rendering pipelines, which are detailed in Sec.~\ref{sec:preliminaries:mlp}, but adding a low-rank decomposed indexing step right before an MLP step. Inside the newly introduced \underline{low-rank decomposed indexing} step, the coordinates of the sampled points along the ray passing through the target pixel (e.g., the green dot in Fig.~\ref{fig:low_rank_decomp_grid_pipeline}) are used to locate the projection of the sampled points in low-rank grids (e.g., the blue, red, and yellow planes in Fig.~\ref{fig:low_rank_decomp_grid_pipeline}), instead of being sent to an MLP directly. Then, the features (e.g., the yellow/red/blue bars in Fig.~\ref{fig:low_rank_decomp_grid_pipeline}) that contain the RGB color and density information of the aforementioned sampled points are interpolated based on the projection coordinates in the aforementioned low-rank grids. Such features from different low-rank grids are then integrated together to be sent to an MLP as the inputs. The remaining steps are identical to the MLP-based rendering pipelines introduced in Sec.~\ref{sec:preliminaries:mlp}.

\textbf{Dominant Scene Representation.} As compared to MLP-based rendering pipelines in Sec.~\ref{sec:preliminaries:mlp}, low-rank-decomposed-grid-based rendering pipelines newly add a set of low-rank decomposed grids as the dominant scene representation. The aforementioned low-rank decomposed grids are usually 2D grids that represent the projection of a 3D grid onto specific 2D spaces, e.g., $xy$, $yz$, $xz$ planes in a $xyz$ 3D coordinate system\cite{reiser2023merf}, and thus can be regarded as performing low-rank decomposition of 3D grids~\cite{reiser2023merf, chen2022tensorf, chen2023factor}. It is worth noting that 3D grids without any decomposition are not scalable for large-scale scenes due to their significant cubic memory requirements, even making the 3D grids sparse, as pointed out in~\cite{reiser2023merf}. Thus, 2D low-rank decomposed grids are better than 3D grids in terms of rendering quality and rendering efficiency trade-offs.

\textbf{Rendering Technique.} Low-rank-decomposed-grid-based rendering pipelines utilize the same volume rendering technique as MLP-based rendering pipelines detailed in Sec.~\ref{sec:preliminaries:mlp}.

\subsection{Hash-Grid-Based Rendering Pipelines}
\label{sec:preliminaries:hash_grid}
\begin{figure}[b]
  \centering
  \vspace{-1.5em}
  \includegraphics[width=1\linewidth]{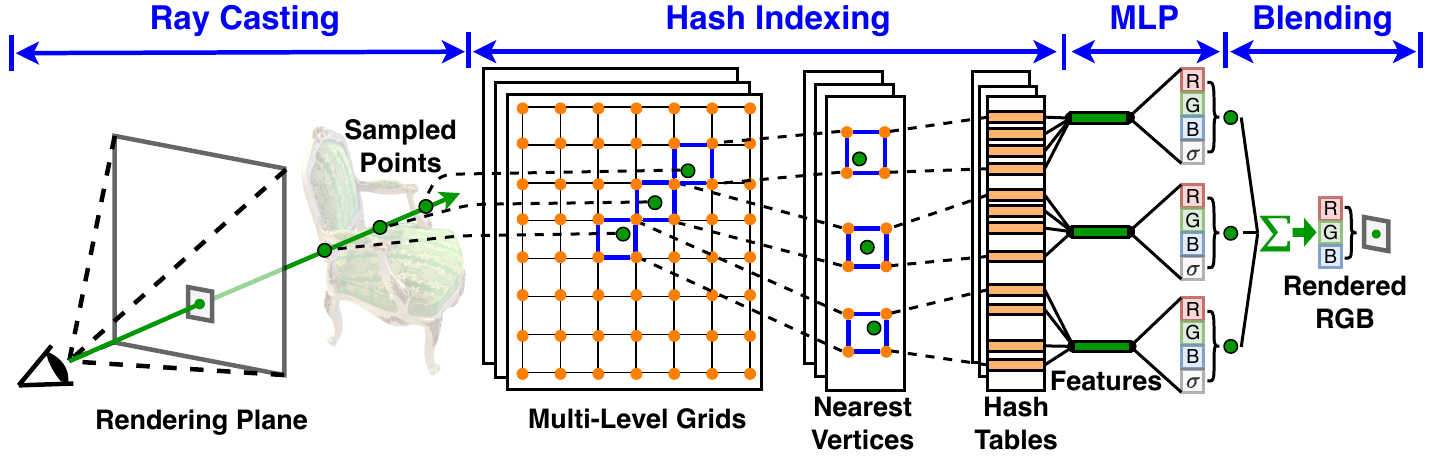}
   \vspace{-2.0em}
\caption{An overview of \textbf{hash-grid-based} rendering pipelines. The detailed description of each step inside the pipelines is in Sec.~\ref{sec:preliminaries:hash_grid}.}
\label{fig:hash_grid_pipeline}
\end{figure}

\textbf{Pipeline Overview.} As illustrated in Fig.~\ref{fig:hash_grid_pipeline}, hash-grid-based pipelines adopt a hash indexing step to replace the low-rank decomposed indexing step of the low-rank-decomposed-grid pipelines detailed in Sec.~\ref{sec:preliminaries:low_rank}. Inside the \underline{hash indexing} step, the coordinates of the sampled points along the ray passing through the target, marked as a green dot in Fig.~\ref{fig:hash_grid_pipeline}, are used to locate the sampled points' nearest vertices in a set of multi-level grids with different grid resolutions. Then, the features that contain the color and density information of each located vertice are fetched from a 1D hash table with a pre-defined random hash mapping function~\cite{instant-ngp}. Finally, the corresponding features of the sampled points are linearly interpolated from the aforementioned nearest vertices. The following steps are the same as the low-rank-decomposed-grid pipelines shown in Sec.~\ref{sec:preliminaries:low_rank}.

\textbf{Dominant Scene Representation.} The key difference between the aforementioned hash-grid-based rendering pipelines and the low-rank-decomposed-grid-based rendering pipelines in Sec.~\ref{sec:preliminaries:low_rank} is adopting hash grids instead of low-rank decomposed grids as the dominant scene representations. The hash grids used in the aforementioned pipelines consist of a set of multi-level 3D grids stored in 1D hash table format. The coordinates of each vertex in the 3D grids can be mapped to a specific index of the hash table via a pre-defined hash mapping function~\cite{instant-ngp}. Since hash collision is allowed in such a design, multiple vertices can be mapped to the same index of the hash table. Thus, hash grids can be regarded as 3D grids with vector quantization~\cite{gray1984vector} by storing the codebook vectors or centroids in the 1D hash table, which also explains why hash-grid-based rendering pipelines are the second most storage efficient ones, as shown in Tab.~\ref{tab:rendering_pipelines}.

\textbf{Rendering Technique.} Hash-grid-based rendering pipelines utilize the same volume rendering technique as the above MLP-based rendering pipelines, as detailed in Sec.~\ref{sec:preliminaries:mlp}.

\subsection{3D-Gaussian-Based Rendering Pipelines}
\label{sec:preliminaries:3d_gaussain}

\begin{figure}[t]
  \centering
  \includegraphics[width=1\linewidth]{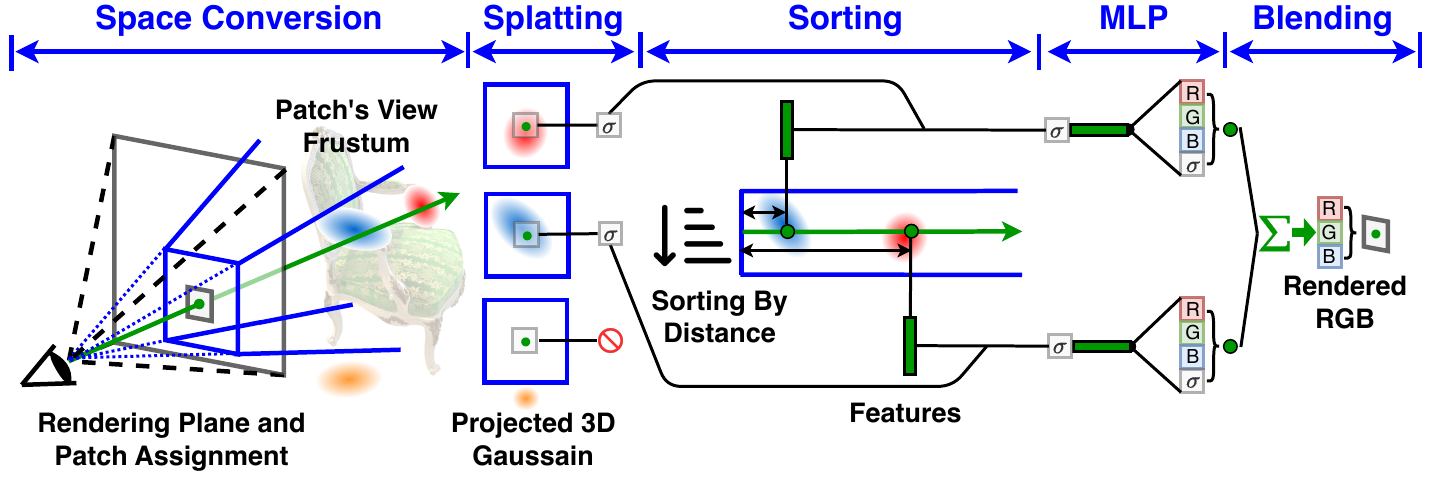}
   \vspace{-2em}
\caption{An overview of \textbf{3D-gaussian}-based rendering pipelines. The detailed description of each step inside the pipelines is in Sec.~\ref{sec:preliminaries:3d_gaussain}.}
\vspace{-1.4em}
\label{fig:3d_gaussain_pipeline}
\end{figure}

\textbf{Pipeline Overview.} As depicted in Fig.~\ref{fig:3d_gaussain_pipeline}, 3D-gaussian-based rendering pipelines can be regarded as a mixture of mesh-based rendering pipelines in Sec.~\ref{sec:preliminaries:mesh} and MLP-based rendering pipelines in Sec.~\ref{sec:preliminaries:mlp}. In other words, 3D-gaussian-based rendering pipelines adopt the same space conversion step as mesh-based rendering pipelines and the same MLP and blending steps as MLP-based rendering pipelines. The function of the newly added splatting and sorting steps are summarized below. In the \underline{splatting} step, the mean and covariance of each 3D gaussian~\cite{kerbl20233d} is used to determine the density of the inspection point between this 3D gaussian and the ray passing through the target pixel to be rendered (e.g., the green dots in Fig.~\ref{fig:3d_gaussain_pipeline}). If the determined density is lower than a pre-defined threshold (e.g., the yellow 3D gaussian in Fig.~\ref{fig:3d_gaussain_pipeline}), then the corresponding gaussian is bypassed for the following steps. Otherwise, the corresponding gaussian is sent to a sorting step as input. Inside the \underline{sorting} step, the gaussian splats that pass the aforementioned thresholding process are sorted by the distance between the rendering plane and the ray-gaussian insertion points. The remaining steps are identical to the MLP-based rendering pipelines introduced in Sec.~\ref{sec:preliminaries:mlp}, and the final blending step follows the gaussian order sorted above to compute the rendered RGB color. 
Although the original 3D Gaussian Spatting (3DGS)~\cite{kerbl20233d} pipeline does not include a neural network, we considered it as a typical neural rendering pipeline in Tab.~\ref{tab:rendering_pipelines}, following the categorization in~\cite{nerfs_on_earth}.
It is worth noting that the MLP in Fig.~\ref{fig:3d_gaussain_pipeline} represents the process of obtaining view-dependent colors based on Spherical Harmonic~\cite{kazhdan2003rotation} coefficients and view directions~\cite{kerbl20233d,luiten2023dynamic,yang2023deformable,keselman2023flexible}, as it can be executed as the vector-matrix multiplication process of MLPs. Thus, the rendering pipeline in our experiment is exactly the original one.

\textbf{Dominant Scene Representation.} 3D gaussians, the dominant scene representation in 3D-gaussian-based rendering pipelines, can be regarded as points with shape and color. Each 3D gaussian consists of (1) the coordinates of the gaussian's centroid/mean, (2) the covariance of the gaussian to presents its shape, (3) the density value of the gaussian, and (4) the RGB color or implicit features of the gaussian. Thus, 3D gaussians use the same file format as point clouds~\cite{ply_file_format} and have similar storage efficiency as compared to mesh-based rendering pipelines, because both of them store explicit discrete approximations of surfaces or semi-transparent volumes.

\textbf{Rendering Technique.} The key rendering technique in 3D-gaussian-based rendering pipelines is splat-based rasterization~\cite{kerbl20233d}, corresponding to the splatting and sorting steps shown in Fig.~\ref{fig:3d_gaussain_pipeline}. Since the sorting step is performed for each patch (e.g., 16$\times$16 pixels as in~\cite{kerbl20233d}) rather than for each target pixel, the sorting cost can be averaged across all pixels in the patch. This patch-based approach also enables real-time rendering on high-end GPUs, such as the RTX 3090~\cite{rtx3090}, even though sorting is more costly than buffering the topmost layer, as compared with mesh-based rendering pipelines.

%% file: sections/3-Benchmark.tex
\section{Motivating Benchmarking}
\label{sec:benchmark}

Based on the detailed analysis of the five typical rendering pipelines in Sec.~\ref{sec:preliminaries} and their summary in Tab.~\ref{tab:rendering_pipelines}, we conclude that \textbf{no single neural rendering pipeline is universally superior in all respects}. Therefore, supporting all five typical pipelines is imperative to ensure compatibility across diverse applications. This conclusion motivates us to conduct a comprehensive benchmark of all five typical neural rendering pipelines across various devices to identify the gap between the current capabilities of neural rendering hardware and the real-time rendering speeds required for diverse applications.

\begin{figure}[t]
  \centering
  \includegraphics[width=1\linewidth]{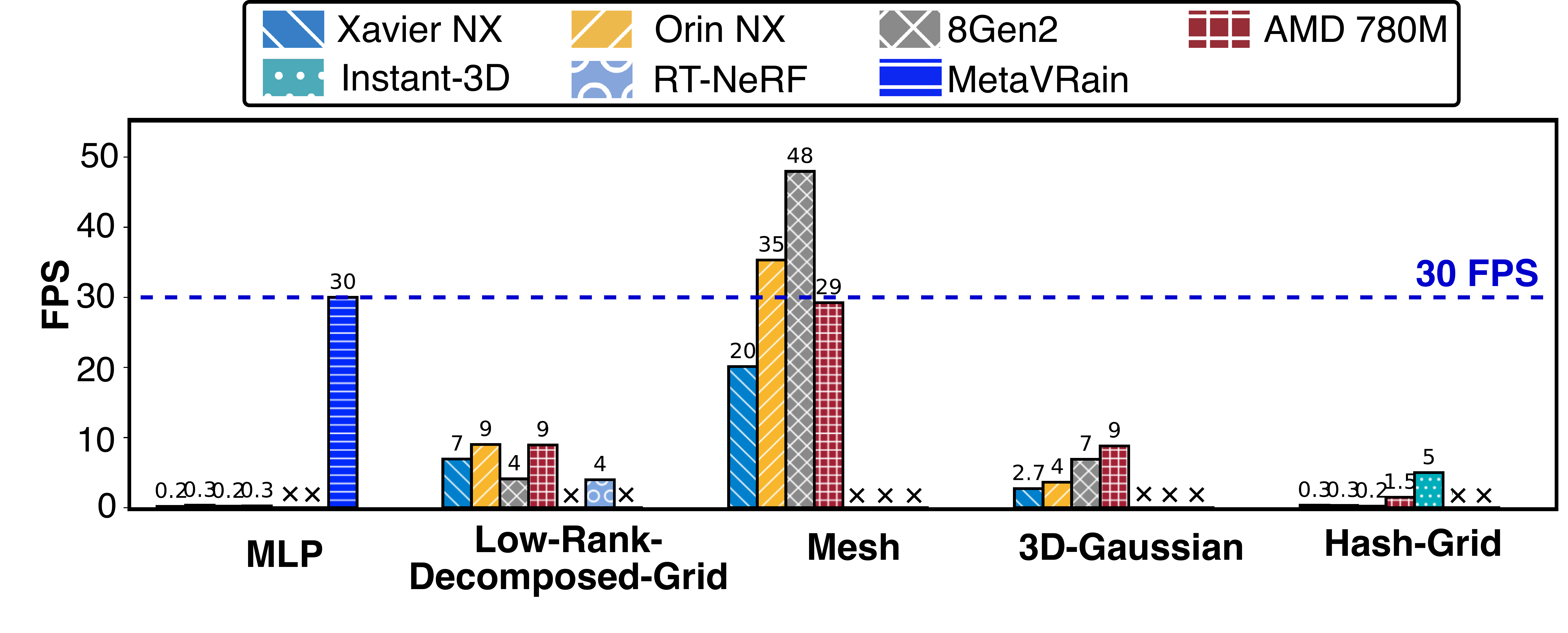}
\vspace{-2em}
\caption{Benchmark on the rendering speed (FPS) of different neural rendering pipelines across different devices/accelerators on the challenging Unbounded-360~\cite{barron2022mip} dataset. The FPS is measured at a rendering resolution of 1280$\times$720, following the settings used in~\cite{reiser2023merf,li2024mixrt}. Bars marked with an ``$\times$'' represent cases of pipelines that are unsupported by the corresponding accelerators.}
\label{fig:benchmark_speed}
\vspace{-1em}
\end{figure}

\subsection{Benchmark Settings}
\label{sec:benchmark:settings}
\textbf{Devices.} We consider four commercial hardware devices and three neural rendering accelerators as the target edge devices. In particular, Qualcomm 8Gen2 mobile development kits~\cite{qualcomm-8g2} is selected as the representative of mobile phones with a typical power consumption of 10 W; NVIDIA Xavier NX~\cite{xnx} and Orin NX~\cite{onx} are chosen as the representatives of edge GPUs with a typical power consumption of 20 W; an x86 desktop with an integrated AMD 780M GPU~\cite{amdgpu} is choose as the representative of desktop with a typical power consumption of 20 W. Meanwhile, Instant-3D~\cite{li2023instant}, RT-NeRF~\cite{li2022rt}, and MetaVRain~\cite{han2023metavrain} are considered in the proposed benchmark as the dedicated neural rendering accelerators for hash-grid-based, low-rank-decomposed-grid-based, and MLP-based rendering pipelines, respectively.

\textbf{Datasets.} To evaluate the FPS achieved by different pipelines across the aforementioned devices, we conduct the benchmark experiments on the challenging Unbounded-360~\cite{barron2022mip} dataset, which is widely used in neural rendering works that target faster rendering speed ~\cite{reiser2023merf,li2024mixrt,yariv2023bakedsdf,rakotosaona2024nerfmeshing}.

\textbf{Implementation Details.} For each typical rendering pipeline illustrated in Sec.~\ref{sec:preliminaries}, we select implementations that represent the state-of-the-art in accuracy vs. efficiency trade-offs. MobileNeRF\cite{mobilenerf}, KiloNeRF~\cite{kilonerf}, MeRF~\cite{reiser2023merf}, Instant-NGP~\cite{instant-ngp}, and 3DGS~\cite{kerbl20233d} have been chosen as representative implementations for mesh-based, MLP-based, low-rank-decomposed-grid-based, hash-grid, and 3D-gaussian-based rendering pipelines, respectively.

\subsection{Benchmark Results and Analysis}
Fig.~\ref{fig:benchmark_speed} shows: none of the existing devices or accelerators \textbf{consistently} achieve a real-time rendering speed of 30 FPS. Across all evaluated settings, only three met the real-time requirements; the others either suffered from inefficiencies or failed on the corresponding device or accelerator due to unsupported operations. These results underscore the urgent need to narrow the gap between the demand for real-time rendering and the current levels of rendering efficiency achieved.

%% file: sections/4-Method.tex
\begin{table*}[!t]
\caption{A summary of (1) how steps from varied pipelines can be clustered into a set of common micro-operators and (2) the corresponding indexing and reduction tasks associated with each unique micro-operator. Here, GEMM stands for General Matrix Multiply.}
\centering
\setlength{\tabcolsep}{2pt}
  \resizebox{1.0\linewidth}{!}
  {
    \begin{tabular}{c||c||c|c|c|c}
    \toprule
     \multirow{2}{*}{\textbf{Micro-Operators}} & \multirow{2}{*}{\textbf{Steps in Typical Pipelines}}  & \multicolumn{3}{c|}{\textbf{Indexing} Task} & \textbf{Reduction} Task \\
    & & \texttt{\{Item\}} & \texttt{\{Dimension\}} & \texttt{\{Function\}}  & \texttt{\{Mem. Access Pattern\}}  \\
    \midrule
 \textbf{Geometric Processing}  & Rasterization and Splatting & Mesh/Gaussian & 1D & Automatic Counter & Continuous\\
    \textbf{Combined Grid Indexing} & Texture and Hash Indexing &   Features & 1D/2D/3D & Random Hash~\cite{instant-ngp}/Linear Indexing~\cite{akenine2018real} & Discrete \\
    \textbf{Decomposed Grid Indexing} & Low-Rank Decomp. Indexing &  Features & 2D/3D & Linear Indexing~\cite{akenine2018real} & Discrete \\
    \textbf{Sorting} & Sorting &  Sorting Keys & 1D & Automatic Counter  & Continuous \\
     \textbf{GEMM} & Others & Scalars & 1D/2D & Automatic Counter & Continuous/Discrete \\
    \bottomrule
    \end{tabular}
    }
  \label{tab:common_ops}
  \vspace{-1em}
\end{table*}

\begin{figure}[b]
  \centering
  \vspace{-1em}
  \includegraphics[width=1\linewidth]{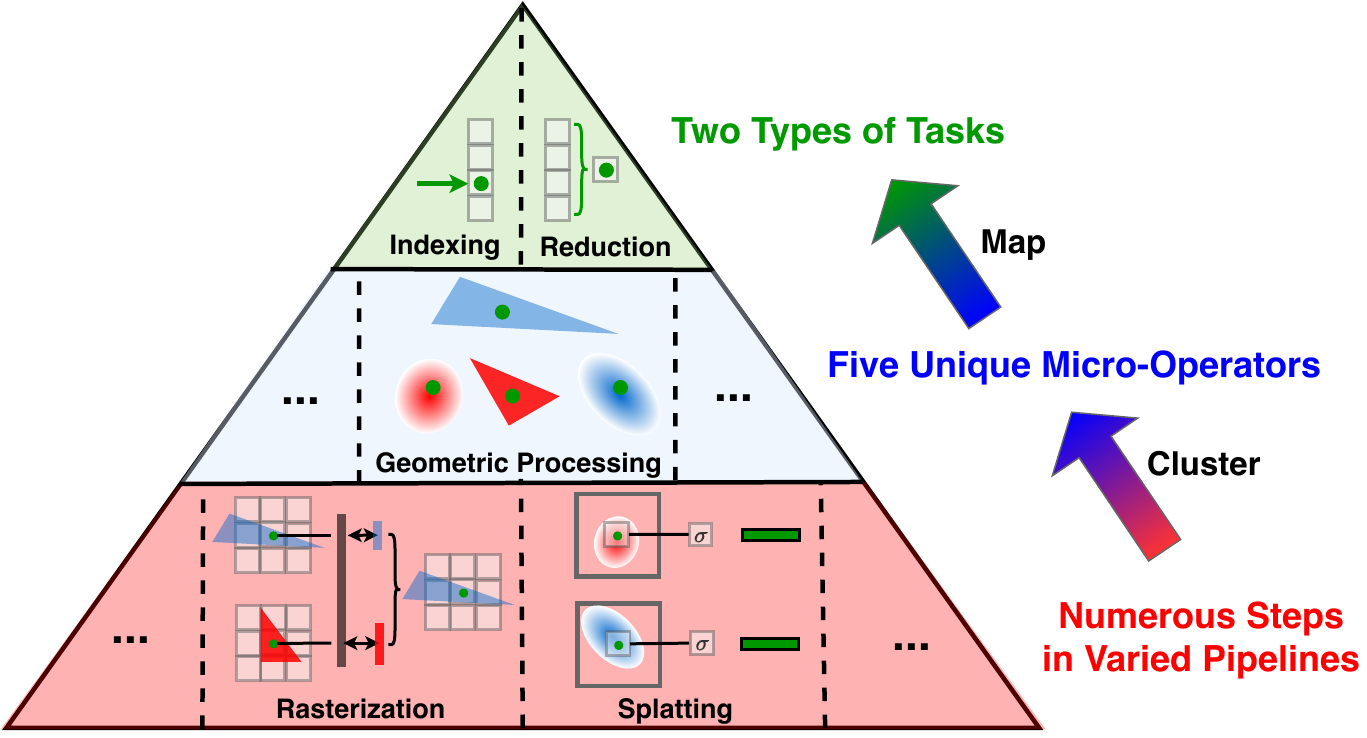}
\vspace{-2em}
\caption{The \textbf{numerous} steps in varied pipelines introduced in Sec.~\ref{sec:preliminaries} can be clustered into \textbf{five} unique micro-operators, each of which can be mapped to the same \textbf{two} types of tasks: one for indexing and one for reduction. This identified uniformity is key to building a unified accelerator for varied neural rendering pipelines.}
\label{fig:micro-operators}
\end{figure}

\section{Common Micro-Operators Across Varied Rendering Pipelines}
\label{sec:ops}

As motivated by the benchmark in Sec.~\ref{sec:benchmark}, we aim to design a unified accelerator to achieve real-time on-device neural rendering. Although the most straightforward method to do so is to merely support a union of all the steps across different pipelines introduced in Sec.~\ref{sec:preliminaries}, i.e., the bottom row in Fig.~\ref{fig:micro-operators}, this approach lacks the flexibility required to support new hybrid pipelines and can suffer from scalability issues. Thus, our goal is to identify a set of common micro-operators, i.e., the middle row in Fig.~\ref{fig:micro-operators}, from the representative rendering pipelines, rather than focusing solely on the steps introduced in Sec.~\ref{sec:preliminaries}. Consequently, our proposed accelerator, built on top of the common micro-operators described below, can support both typical and the latest hybrid pipelines, as illustrated in Sec.~\ref{sec:exp_typical} and Sec.~\ref{sec:exp_hybrid}, respectively.

Among the different steps in the five typical pipelines summarized in Sec.~\ref{sec:preliminaries}, we observe a common pattern: these \textbf{numerous} steps can be clustered into \textbf{five} unique micro-operators, with each corresponding to the same \textbf{two} types of tasks: one for indexing and the other for reduction, as illustrated in Fig.~\ref{fig:micro-operators}. This observation aligns with the findings of early General-Purpose Graphics Processing Unit (GPGPU) compilers~\cite{buck2004brook}, which demonstrated that the computation graphs of various scientific computing applications can also be decomposed into two types of tasks: one for gathering/indexing and the other for reduction. For example, as illustrated in Fig.~\ref{fig:micro-operators} and summarized in Tab.~\ref{tab:common_ops}, both the rasterization step in mesh-based rendering pipelines and the splatting step in 3D-gaussian-based rendering pipelines can be clustered into the Geometric Processing micro-operator. In particular, the Geometric Processing micro-operator involves (1) the \underline{\textbf{indexing}} task that indexes \texttt{\{Mesh/Gaussian\}} from a \texttt{\{1D\}} tensor (e.g., the coordinates of each gaussian centroid), with the index retrieved by \texttt{\{Automatic Counter\}} (i.e., automatically increments the index by one each time the function is called), and (2) the \underline{\textbf{reduction}} task that performs reduction within a set of \texttt{\{Continuous\}} memory addresses (e.g., the memory addresses of gaussian centroids' coordinates that are continuously stored in memory).

Similar to the aforementioned Geometric Processing micro-operator, as shown in Tab.~\ref{tab:common_ops}, all five unique micro-operators can generally be mapped to: (1) \underline{\textbf{Indexing}} task, ``indexing \texttt{\{Item\}} from a \texttt{\{Dimension\}} tensor, with the index retrieved by \texttt{\{Function\}}''; (2) \underline{\textbf{Reduction}} task, ``performing reduction within a set of \texttt{\{Mem. Access Pattern\}} memory addresses''. The aforementioned uniformity provides a compelling reason to design a unified hardware architecture that supports all the typical rendering pipelines discussed in Sec.~\ref{sec:preliminaries}, as well as the latest hybrid pipelines, by accelerating the identified set of common micro-operators.

\begin{figure*}[!t]
  \centering
  \includegraphics[width=1\linewidth]{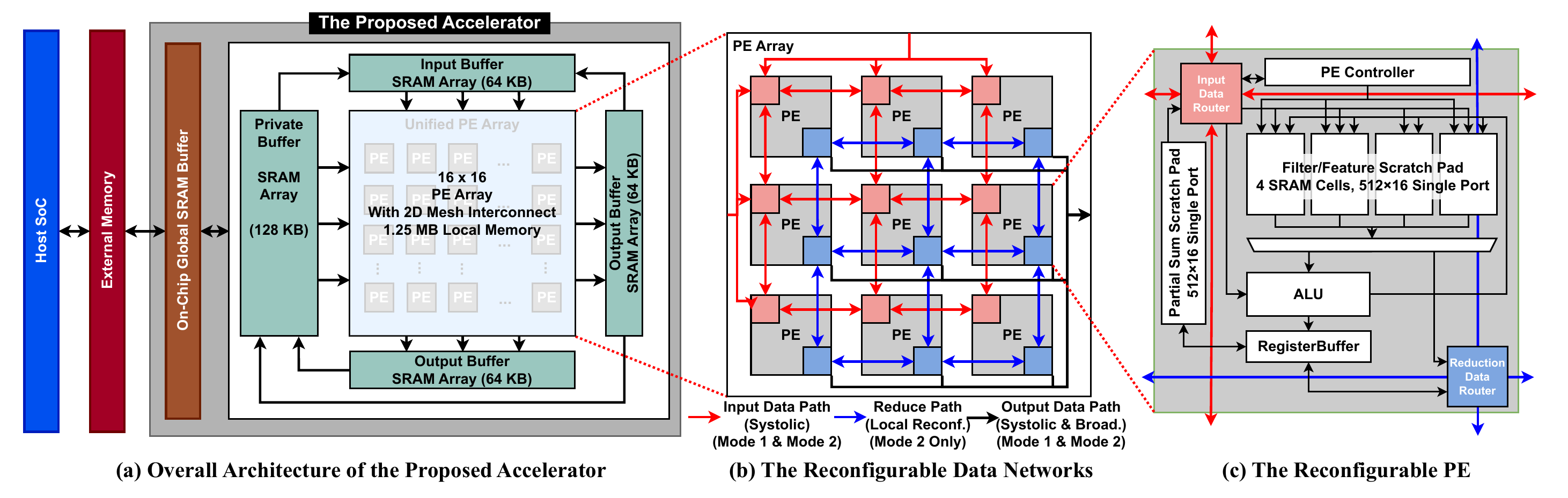}
  \vspace{-2em}
\caption{An overview of the proposed hardware architecture design, detailing (a) the overall architecture of the proposed Uni-Render accelerator for varied rendering pipelines, (b) the reconfigurable input/reduction data networks within the PE array, and (c) the block diagram of each individual reconfigurable PE.
}
  \vspace{-1em}
\label{fig:HW-Overall}
\end{figure*}

\section{Uni-Render Accelerator Design: Hardware Architecture}
\label{sec:arch}

\subsection{Architecture Overview}
\label{sec:arch:overview}

Building upon our insight that numerous steps in various rendering pipelines can be clustered into five unique micro-operators, each corresponding to the same two types of tasks, as summarized in Fig.~\ref{fig:micro-operators}, we propose a reconfigurable hardware architecture to accelerate these micro-operators, thereby supporting both typical and the latest hybrid rendering pipelines. As depicted in Fig.~\ref{fig:HW-Overall} (a), the entire system comprises (1) the proposed accelerator, (2) a host System on Chip (SoC) to issue commands and execute certain lightweight computations, and (3) an external memory designed for storing intermediate results.

The key design elements in the proposed accelerator that enable the desired reconfigurability for different micro-operators are (1) the reconfigurable input/reduction data networks, as shown in Fig.~\ref{fig:HW-Overall} (b), designed to execute the \textbf{reduction} tasks, and (2) the reconfigurable Processing Elements (PEs), as demonstrated in Fig.~\ref{fig:HW-Overall} (c), designed to execute the \textbf{indexing} tasks. We detail the design of the reconfigurable input/reduction data networks and the reconfigurable PE in Sec.~\ref{sec:arch:reconfigurable_data_networks} and Sec.~\ref{sec:arch:reconfigurable_PE}, respectively.

\subsection{Reconfigurable Input/Reduction Data Networks}
\label{sec:arch:reconfigurable_data_networks}
The configuration of the reconfigurable input/reduction data networks within the PE array is determined by the status of the (1) \textbf{data paths}, represented by red/blue arrows in Fig.~\ref{fig:HW-Overall} (b), and (2) the \textbf{data routers} denoted by red/blue boxes in Fig.~\ref{fig:HW-Overall} (c). Additionally, the ``output data path'' represented by black arrows in Fig.~\ref{fig:HW-Overall} is the data path used by PEs to write results to the output buffer. In particular, each PE in the array is interconnected through these input/reduction data networks, enabling the configuration of the entire PE array into two distinct modes. In \textit{Mode 1}, the PE array functions in a ``\textbf{systolic-array-like}'' way to pass data between PEs~\cite{tpu}, enhancing efficiency for the GEMM micro-operator crucial in multiple rendering pipelines as detailed in Sec.~\ref{sec:ops}. In \textit{Mode 2}, the PE array is reconfigured as a \textbf{pipeline}, dynamically adapting the input/reduction data networks to suit the memory access patterns required for the \textbf{reduction} tasks.

\subsection{Reconfigurable PEs}
\label{sec:arch:reconfigurable_PE}

Since indexing tasks are more complex than reduction tasks, as summarized in Tab.~\ref{tab:common_ops}, the reconfigurable PE for \textbf{indexing} requires more status types for configuration compared to the reconfigurable data networks for reduction tasks.
As illustrated in Fig.~\ref{fig:HW-Overall} (c), the configuration of each PE is determined by the status of the following four modules within it: (1) \textbf{PE controller}, which regulates the dataflow within the PE and generates addresses for the Filter/Feature (FF) scratch pad; (2) \textbf{FF scratch pad}, consisting of four SRAM cells, each sized 512$\times$16 bits. This scratch pad stores the detailed description of the target indexing tasks. The address to access this scratch pad can come from the aforementioned PE controller, which integrates a simple counter for automatic counter-based indexing, or from the Arithmetic Logic Unit (ALU) described later; (3) \textbf{ALU}, which includes four INT16 MACs (mainly for index computations to execute indexing tasks), four BF16 MACs (mainly for feature computations based on the indices retrieved from indexing tasks), and four special function units. These MACs can be configured into different layouts to accommodate the indexing tasks required by different micro-operators. For example, they can be configured as an Adder Tree for the indexing tasks in the GEMM and Combined/Decomposed Grid Indexing micro-operators~\cite{huang2022high}; (4) \textbf{Partial Sum (PS) scratch pad}, which is connected to a register buffer and serves as the output memory.

\begin{table*}[!t]
\caption{A summary of the status of the hardware modules in the proposed reconfigurable input/reduction data networks and reconfigurable PEs for supporting different micro-operators.}
\centering
\setlength{\tabcolsep}{6pt}
  \resizebox{1.0\linewidth}{!}
  {
    \begin{tabular}{c||c|c||c|c|c|c}
    \toprule
     \multirow{2}{*}{\textbf{Micro-Operators}} & \textbf{Input Data} & \textbf{Reduction Data} & \multirow{2}{*}{\textbf{PE Controller}} & \multirow{2}{*}{\textbf{FF Scratch Pad}} & \multirow{2}{*}{\textbf{ALU}} & \multirow{2}{*}{\textbf{PS Scratch Pad}} \\
      & \textbf{Paths\&Routers} & \textbf{Paths\&Routers} &  & &  & \\
    \midrule
 \textbf{Geometric Processing}  & Off & Off & Rasterization Control & Geometry Representation & Vector Mode & Z-Buffer\\
    \textbf{Combined Grid Indexing} & On & Horizontally On& Grid Control & Grid Features & Index Function & Off \\
    \textbf{Decomposed Grid Indexing} & On & Fully On & Grid Control & Grid Features & Index Function & Off  \\
    \textbf{Sorting} & Off & Off & Sorting Control & Sorting Elements & Comparator & Off \\
     \textbf{GEMM} & On & Off & GEMM Control & Model Weights & Adder Tree Mode & Output Features \\
    \bottomrule
    \end{tabular}
    }
  \label{tab:dataflows}
  \vspace{-1.5em}
\end{table*}

\section{Uni-Render Accelerator Design:\\Dataflows for the Micro-Operators}
\label{sec:dataflow}

By configuring the status of the \textbf{data paths} and \textbf{data routers} in the reconfigurable data networks, as well as the \textbf{PE controller}, \textbf{FF scratch pad}, \textbf{ALU}, and \textbf{PS scratch pad} in each reconfigurable PE, we can map the common micro-operators, shared by varied rendering pipelines, to the designed reconfigurable architecture detailed in Sec.~\ref{sec:arch}. During the aforementioned mapping process, the summary in Tab.~\ref{tab:dataflows} is referred to configure the status of each aforementioned hardware module. Below, we elaborate on the dataflow for each micro-operator.

\begin{figure}[t]
  \centering
  \includegraphics[width=1\linewidth]{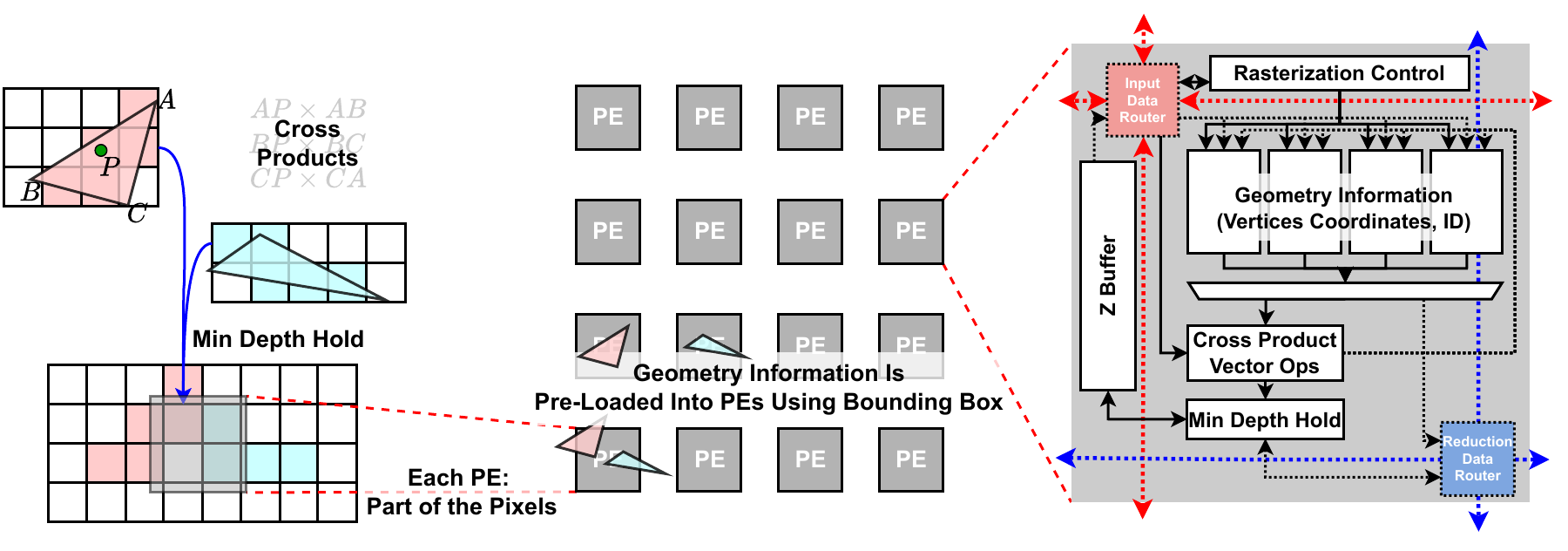}
\vspace{-2em}
\caption{The dataflow of the \textbf{Geometric Processing} micro-operator on the proposed reconfigurable architecture.}
    \vspace{-1.5em}
\label{fig:GeometryProcessing}
\end{figure}

\textbf{Dataflow for \underline{Geometric Processing}.}
To leverage the fact that the Geometric Processing micro-operator can be naturally parallelized across different pixels~\cite{laine2011high,kerbl20233d}, we \textbf{map} a specific region of pixels to each PE, as depicted in Fig.~\ref{fig:GeometryProcessing}. Based on the aforementioned mapping, when executing the \textbf{indexing} task within this micro-operator, a built-in automatic counter in the PE controller sequentially accumulates memory addresses, which are sent to the FF scratch pad for geometry data retrieval. Once all the required data is obtained, the ALU performs barycenter~\cite{laine2011high} calculations to determine if the pixel is inside the corresponding geometry primitive (e.g., triangle). Concurrently, depth values are computed based on the results of these barycenter calculations.
For the \textbf{reduction} task, the Geometric Processing micro-operator incorporates a minimum depth holding mechanism within each PE, where the index of the nearest geometry primitive is preserved. Upon completing the tasks for one pixel, the corresponding index and barycenter coefficients of the nearest geometry primitive are stored in the PS scratch pad (equivalent to the Z-Buffer in traditional rasterization engines~\cite{laine2011high}). The results from the executed micro-operator are read out once the PE has completed the aforementioned tasks.

\begin{figure}[b]
  \centering
  \vspace{-1.5em}
  \includegraphics[width=1\linewidth]{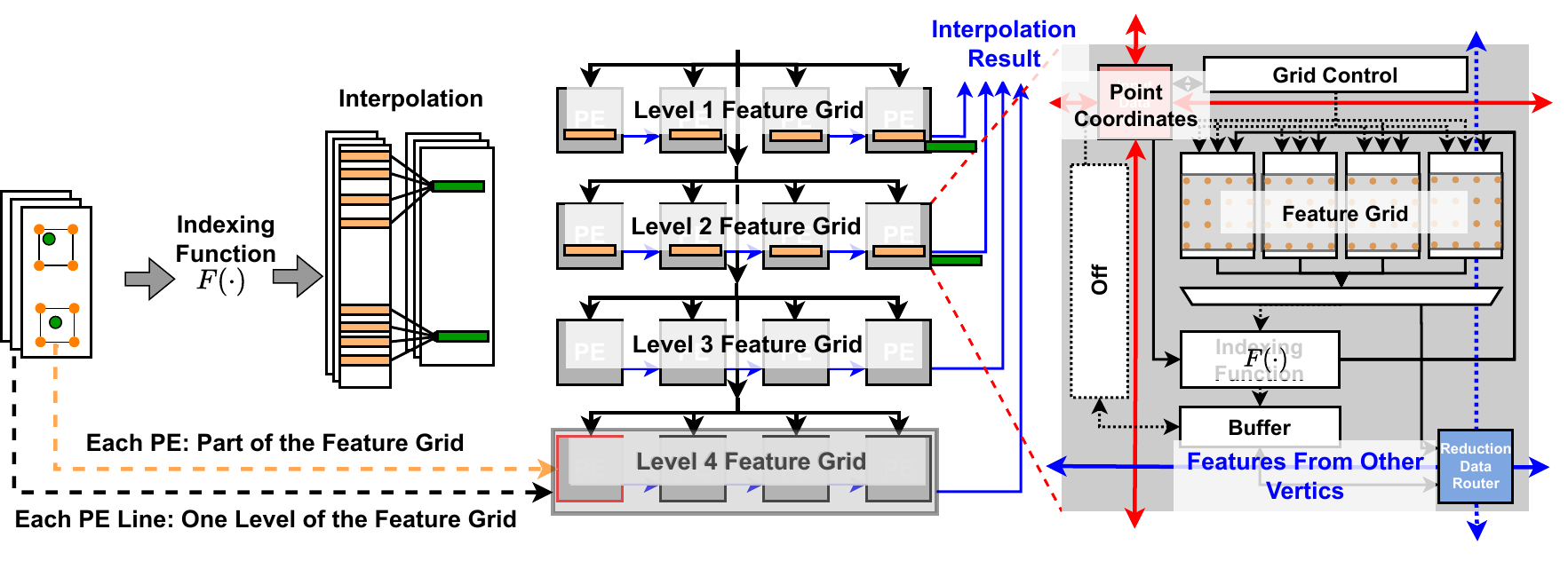}
\vspace{-2em}
\caption{The dataflow of the \textbf{Combined Grid Indexing} micro-operator on the proposed reconfigurable architecture.}
\label{fig:CombinedGridIndexing}
\end{figure}

\begin{figure}[t]
  \centering
  \includegraphics[width=1\linewidth]{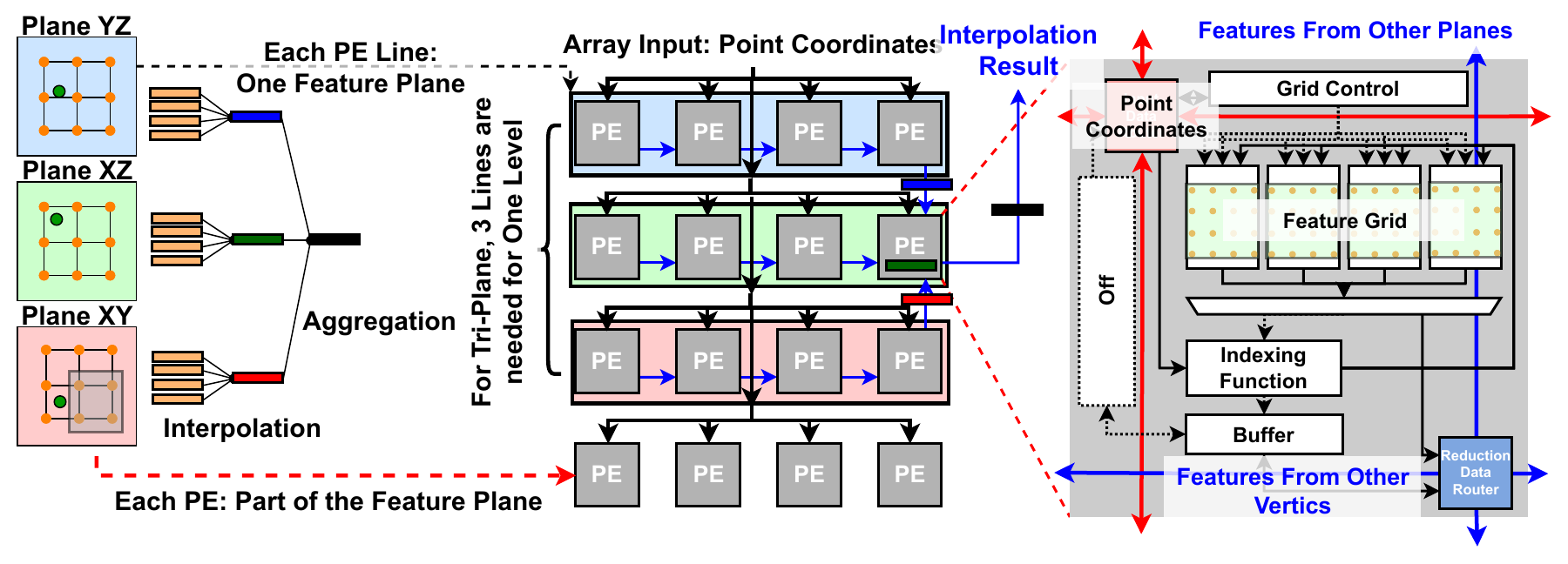}
\vspace{-2em}
\caption{The dataflow of the \textbf{Decomposed Grid Indexing} micro-operator on the proposed reconfigurable architecture.}
    \vspace{-1.5em}
\label{fig:DecomposedGridIndexing}
\end{figure}

\textbf{Dataflow for \underline{Combined Grid Indexing}.}
To fulfill the need for bilinear or trilinear interpolation required by the Combined Grid Indexing micro-operator, and to leverage the layout of the feature grid across multiple levels~\cite{instant-ngp}, we \textbf{map} the feature grid in this micro-operator to the PE array in a 2D layout, as shown in Fig.~\ref{fig:CombinedGridIndexing}. In particular, each PE line corresponds to a level of features, with different PEs in the same line representing the interpolation candidates, i.e., parts of the feature grid. To execute the \textbf{indexing} task within the aforementioned mapping design, the ALUs are configured to compute addresses of features to be retrieved from the FF Scratch Pad. This configuration accommodates the more complex indexing computations of this micro-operator, as compared to those in the Geometric Processing micro-operator. For the \textbf{reduction} task, the reduction data network is activated horizontally and configured as a weighted adder tree to interpolate the retrieved features, aligning with the mapping that executes interpolation within each PE line instead of across multiple PE lines.

\textbf{Dataflow for \underline{Decomposed Grid Indexing}.}
Since the feature grid in the Decomposed Grid Indexing micro-operator consists of multiple feature planes and requires both interpolation within each plane and aggregation across different planes, we \textbf{map} each plane of the feature grid to one PE line for this micro-operator. Similar to the mapping for the Combined Grid Indexing micro-operator, different PEs in the same line correspond to the interpolation candidates from the same plane. For the \textbf{indexing} task, we use the same setup of the configurable hardware modules as those in the Combined Grid Indexing micro-operator, as indicated in Tab.~\ref{tab:dataflows}. However, for the \textbf{reduction} task, because the interpolated features from different planes need to be aggregated together, the reduction data network is fully (both horizontally and vertically) activated. In particular, the reduction data network first performs weighted addition (interpolation) within each PE line based on the features obtained from the same feature plane, and then performs weighted multiplication (aggregation) across multiple PE lines based on the features from different planes.

\begin{figure}[b]
  \centering
    \vspace{-1.5em}
  \includegraphics[width=1\linewidth]{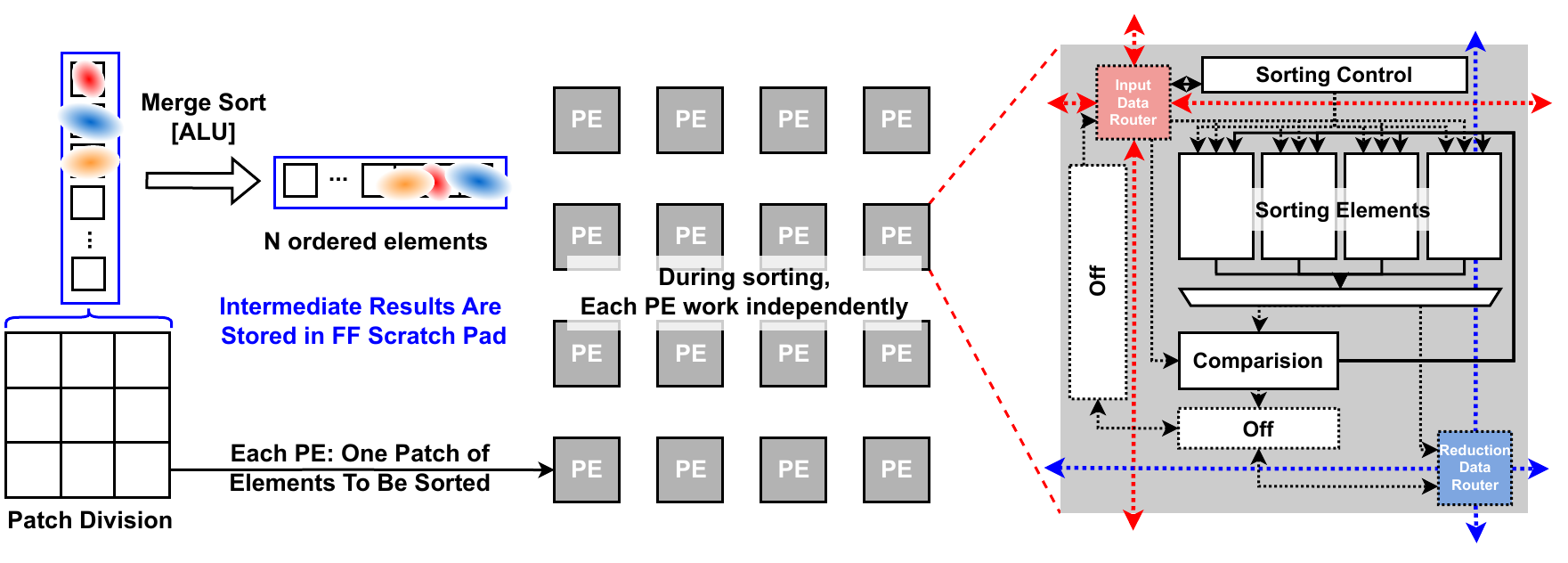}
\vspace{-2em}
\caption{The dataflow of the \textbf{Sorting} micro-operator on the proposed reconfigurable architecture.}
\label{fig:Sorting}
\end{figure}

\textbf{Dataflow for \underline{Sorting}.}
Similar to the parallelism in the Geometric Processing micro-operator, the Sorting micro-operator can be parallelized across the divided patches of the image to be rendered. This is possible because pixels within the same patch usually share the same rendering order, as observed by~\cite{kerbl20233d}. Thus, we \textbf{map} the set of unordered elements within the same patch to each PE. To execute the \textbf{indexing} task, a counter in the PE controller is used to select the target elements for sorting. To avoid complex unordered indexing, we adopt a merge sort implementation~\cite{merge_sort}, which gradually merges smaller ordered sets into larger ones. Specifically, the unsorted elements are pre-loaded into the FF scratch pad, and the ALU is reconfigured to act as comparators that determine the order of elements from these smaller sets. The smaller sets, once ordered, are written back to the FF scratch pad iteratively until the sorting within the target patch is completed. For the \textbf{reduction} task, since each PE operates independently in this mapping, the reduction data network is deactivated.

\begin{figure}[t]
  \centering
  \includegraphics[width=1\linewidth]{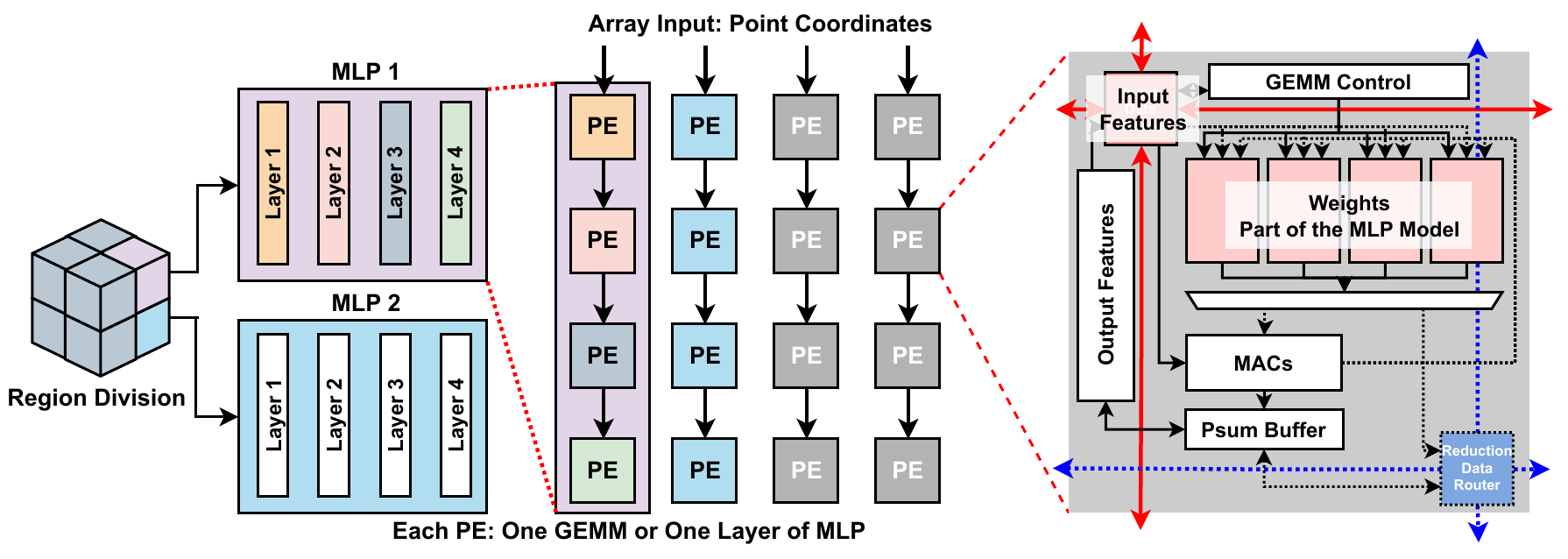}
\vspace{-2em}
\caption{The dataflow of the \textbf{GEMM} micro-operator on the proposed reconfigurable architecture.}
\label{fig:GEMM}
    \vspace{-1.1em}
\end{figure}

\textbf{Dataflow for \underline{GEMM}.}
Motivated by the observation that MLPs used in neural rendering pipelines typically have a small model size (e.g., fewer than 1 million parameters in NeRF~\cite{mildenhall2020nerf} compared to over 100 million parameters in VGG16~\cite{simonyan2014very}), but operate with significantly larger batch sizes (e.g., greater than 1024 for NeRF~\cite{mildenhall2020nerf} compared to a typical batch size of 1 for VGG16~\cite{simonyan2014very}), we leverage this high batch size and small model size characteristic of MLPs in neural graphics, distinct from prior deep learning accelerators~\cite{eyeriss,shidiannao}, to \textbf{map} the GEMM micro-operator, primarily invoked by MLPs in various rendering pipelines, to our proposed reconfigurable architecture in a weight-stationary manner, as shown in Fig.~\ref{fig:GEMM}.
The aforementioned weight-stationary manner is selected to maximize data reuse and minimize the required local memory size for workloads with a relatively large batch size and small model size. For the \textbf{indexing} task, we use a counter inside the PE controller instead of the more complex ALU setups adopted by Combined Grid Indexing micro-operator, because of the regular and dense memory access pattern in the GEMM micro-operator. Moreover, the FF Scratch Pad is employed for storing MLP weights, and the ALU performs standard multiplication-accumulation operations. For the \textbf{reduction} task, it is executed inside each PE, where the partial sum is reduced within the PE and the output of the MLPs is stored in the PS scratch pad. Thus, the reduction data network is deactivated.

%% file: sections/5-Experiments.tex
\section{Experiments}

\subsection{Experiments Settings}
\label{sec:experiments_settings}

\textbf{Datasets \& Baselines.}
\underline{Datasets}: To evaluate the rendering efficiency of our proposed accelerator, following the benchmarking effort in Sec.~\ref{sec:benchmark}, we conduct experiments on the challenging Unbounded-360~\cite{barron2022mip} dataset. \underline{Baselines}: We consider all four edge devices and three accelerators we benchmarked in Sec.~\ref{sec:benchmark} as the baselines and we implement all five typical neural rendering pipelines following the implementation with the fastest rendering speeds, as introduced in Sec.~\ref{sec:benchmark}. Specifically, the four baseline commercial devices adopt the optimized WebGL~\cite{webgl}-based software implementation, and all steps are offloaded to the GPUs in the corresponding devices to execute the target rendering pipelines.
\begin{figure}[t]
    \centering
    \includegraphics[width=\linewidth]{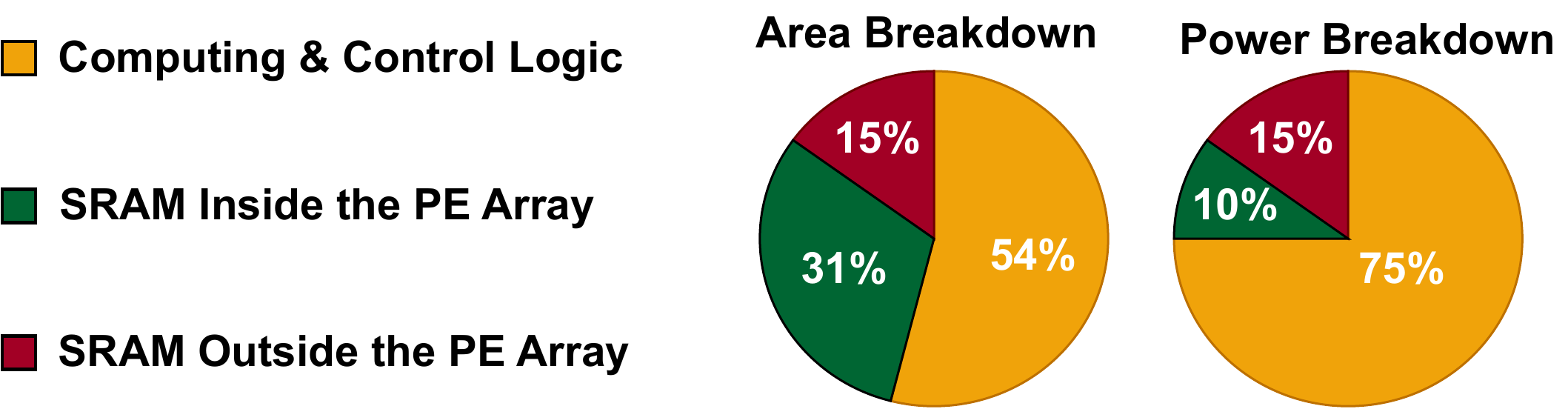}
\vspace{-1.5em}
    \caption{\textcolor{black}{The area and power breakdown of our proposed accelerator.}}
    \vspace{-1.5em}
    \label{fig:breakdown}
\end{figure}
\textbf{Hardware Implementation of Our Proposed Accelerator.}
To evaluate the power and the area of the proposed accelerator, we implement our accelerator in RTL. \textcolor{black}{The RTL is further synthesized, placed, and routed using Synopsys Design Compiler~\cite{synopsys_design_compiler} and Cadence Innovus~\cite{cadence_innovus}, based on a commercial 28 nm CMOS technology. The operation voltage is set to 0.9 V, and the clock frequency is 1GHz with the delay corner set to typical.} 
In all our experiments, we set the size of the on-chip global SRAM buffer to 256 KB. In addition, we develop a cycle-accurate simulator to estimate the efficiency of our proposed accelerator on different rendering pipelines with the assumption of a 59.7 GB/s DRAM bandwidth, which is the same as the typical bandwidth in LPDDR4-1866 used in the baseline hardware device~\cite{xnx}. \textcolor{black}{The proposed accelerator consumes an area of 14.96 mm$^2$ and has a typical power consumption of 5.78 W, which is comparable to the approximately 5 W typically seen in edge devices~\cite{raspi4,qualcomm-8g2}.}
The breakdown of area and power consumption is shown in Fig.~\ref{fig:breakdown}. Following~\cite{li2023instant,sanger,ham20203}, the power estimation excludes DRAM and corresponds to the setting of a 16$\times$16 PE array. 
The rendering speed is measured as the average FPS across all test views and all scenes in the benchmark dataset.

\begin{figure}[!b]
    \centering
    \vspace{-1.em}
    \hspace{-1.5em}
    \includegraphics[width=1.04\linewidth]{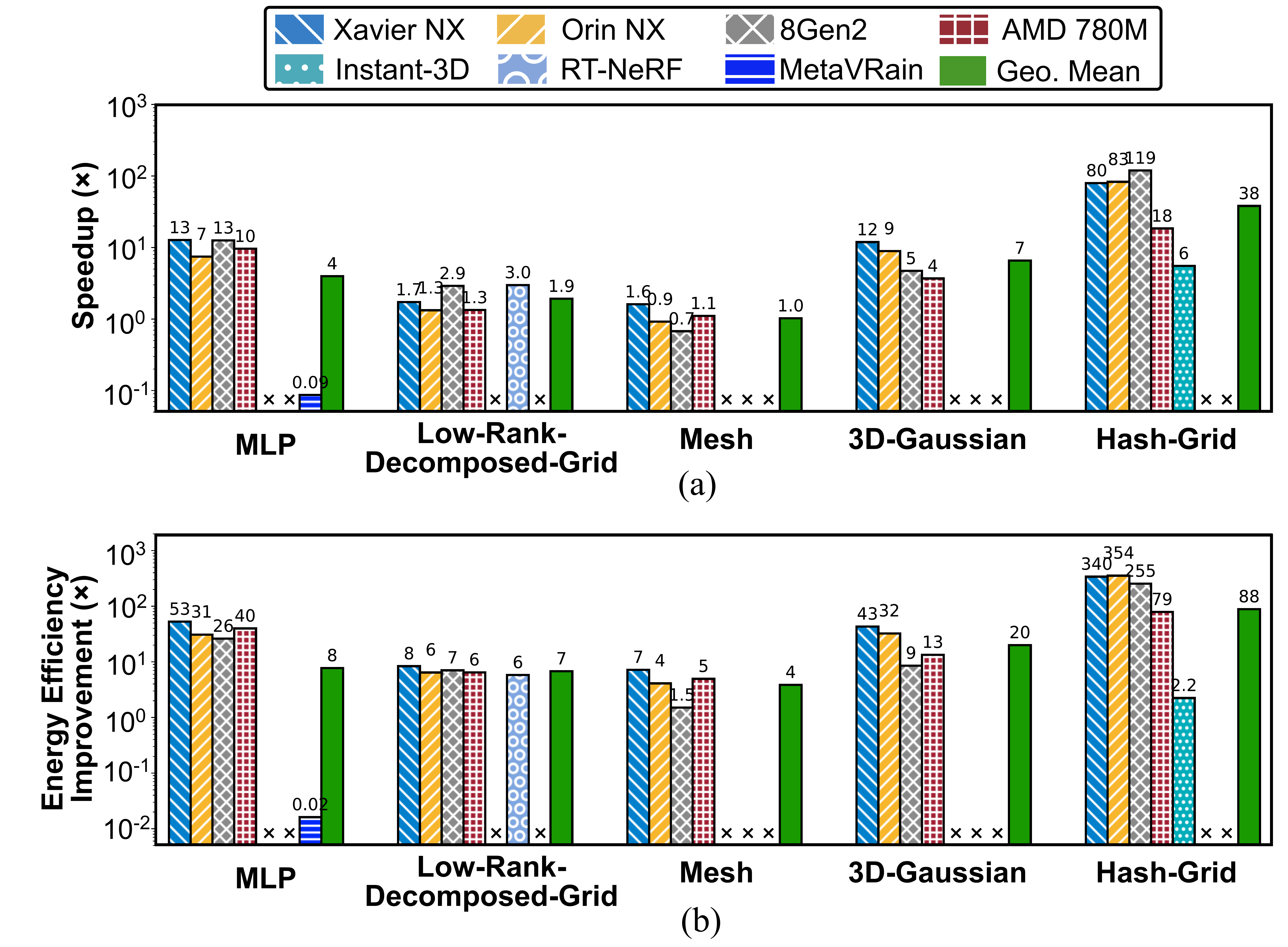}
\vspace{-2em}
    \caption{The (a) speedup and (b) \textcolor{black}{energy efficiency} improvement achieved by our proposed accelerator over the baseline devices/accelerators on five typical rendering pipelines for the scenes from the Unbounded-360~\cite{barron2022mip} dataset. Bars marked with an ``$\times$'' indicate cases of unsupported rendering pipelines in the corresponding baseline accelerators.}
    \label{fig:exp_typical}
\end{figure}

\subsection{Evaluation on \textbf{Typical} Rendering Pipelines}
\label{sec:exp_typical}

To evaluate the effectiveness of our proposed accelerator, we summarize the rendering speedup and energy efficiency improvement it achieves over existing devices and accelerators across different rendering pipelines in Fig.~\ref{fig:exp_typical}. We observe that the proposed accelerator consistently achieves an impressive speedup over existing devices and accelerators. 

\textbf{Comparison with Commercial Devices.}
When compared with the four baseline commercial devices, i.e., 8Gen2~\cite{qualcomm-8g2}, Xavier NX~\cite{xnx}, Orin NX~\cite{onx}, and AMD 780M~\cite{amdgpu}, our proposed accelerator demonstrates a speedup of 0.7$\times$ to 119$\times$ and an energy efficiency improvement of \textcolor{black}{1.5}$\times$ to \textcolor{black}{354}$\times$ across various rendering pipelines.
Specifically, since NVIDIA Edge GPU Orin NX~\cite{onx} and Qualcomm Mobile SoC 8Gen2~\cite{qualcomm-8g2} are highly optimized for mesh-based rendering pipelines, our proposed accelerator achieves only 0.9$\times$ and 0.7$\times$ rendering speeds, respectively, when evaluated on the mesh-based pipeline~\cite{mobilenerf}, due to the reconfiguration overheads as discussed in Sec.~\ref{sec:reconfiguration_overhead}. However, it is worth noting that our proposed accelerator still achieves a \textcolor{black}{4}$\times$ and \textcolor{black}{1.5}$\times$ energy efficiency improvement on the mesh-based pipeline~\cite{mobilenerf}, compared to Orin NX~\cite{onx} and 8Gen2~\cite{qualcomm-8g2}, respectively. The speedup and efficiency improvement of our proposed accelerator over commercial devices result from (1) accelerating all the identified common micro-operators, as illustrated in Sec.~\ref{sec:dataflow}, and (2) the optimized hardware configurations for the neural rendering pipeline, as discussed in Sec.~\ref{sec:abl_PE_SRAM_sizes}.

\textbf{Comparison with Prior Neural Rendering Accelerators.}
For the three baseline neural rendering accelerators, i.e., Instant-3D~\cite{li2023instant}, RT-NeRF~\cite{li2022rt}, and MetaVRain~\cite{han2023metavrain}, they are specially designed for hash-grid-based, low-rank-decomposed-grid-based, and MLP-based rendering pipelines, respectively. Thus, they cannot support other pipelines, which corresponds to the unavailable results represented by bars marked with an ``$\times$'' in Fig.~\ref{fig:exp_typical}. In contrast, our proposed accelerator achieves consistent improvement across all pipelines. The reconfigurability of the proposed accelerator, which flexibly supports varying representative rendering pipelines and the most recent hybrid pipelines, ensures that the proposed accelerator can outperform existing accelerators in terms of the average speedup on multiple rendering pipelines.

Specifically, our proposed accelerator achieves a speedup of 3$\times$ and an energy efficiency improvement of 6$\times$ over RT-NeRF~\cite{li2022rt} on the low-rank-decomposed-grid rendering pipeline~\cite{reiser2023merf}, because RT-NeRF~\cite{li2022rt} is specially designed for sparse 2D grids, while the most commonly-used efficiency low-rank-decomposed-grid rendering pipeline~\cite{reiser2023merf} consists of dense 2D grids and sparse 3D grids. Thanks to the reconfigurability of our proposed accelerator, which is enabled by accelerating the common micro-operators summarized in Tab.~\ref{tab:common_ops}, it can more efficiently execute the low-rank-decomposed-grid rendering pipeline~\cite{reiser2023merf}.
Similarly, our proposed accelerator achieves a speedup of 6$\times$ and an energy efficiency improvement of \textcolor{black}{2.2}$\times$ over Instant-3D~\cite{li2023instant} on the hash-grid rendering pipeline~\cite{reiser2023merf}, because Instant-3D~\cite{li2023instant} is optimized for relatively smaller-scale objects~\cite{mildenhall2020nerf} and bounded indoor scenes~\cite{dai2017scannet,courteaux2022silvr}, and Instant-3D~\cite{li2023instant} cannot be reconfigured for other pipelines or scenes. In contrast, our proposed accelerator can efficiently support the large-scale unbounded scenes~\cite{barron2022mip}, which are more challenging and commonly used in current neural rendering applications~\cite{reiser2023merf,yariv2023bakedsdf}.

However, such reconfigurability can result in lower rendering speeds compared to a specifically designed accelerator for some rendering pipelines. For instance, as compared to MetaVRain~\cite{han2023metavrain}, which is specially designed for MLP-based rendering pipelines, our proposed accelerator only achieves MetaVRain's~\cite{han2023metavrain} 2\% energy efficiency, i.e., 10\% FPS with 5$\times$ more power consumption, for the MLP-based pipeline~\cite{kilonerf}. This efficiency reduction comes from two parts: (1) MetaVRain~\cite{han2023metavrain} adopts Pixel-Reuse~\cite{han2023metavrain} to avoid rendering similar pixels from nearby images, reducing the computation by $\sim$ 20$\times$. However, our proposed accelerator does not adopt Pixel-Reuse~\cite{han2023metavrain} as the default setting because Pixel-Reuse~\cite{han2023metavrain} assumes that the nearby images always have similar corresponding camera poses, which is not true for neural rendering applications with fast camera motion~\cite{klenk2023nerf}; (2) Our proposed accelerator is not solely optimized for the MLP-based rendering pipeline. Thus, the reconfiguration overhead discussed in Sec.~\ref{sec:reconfiguration_overhead} also affects the achieved speedup and energy efficiency.

\begin{table}[!t]\centering
\caption{Our proposed accelerator achieves real-time rendering speeds on typical pipelines. The FPS is measured on the NeRF Synthetic dataset~\cite{mildenhall2020nerf}, following the settings in~\cite{li2022rt,Lee2023Neurex}.}
\scriptsize
\resizebox{\linewidth}{!}{\begin{tabular}{c|c||c|c}\toprule
Rendering & Reference & Rendering & Real-Time \\
Pipeline & Algorithm & Speed & Rendering \\
\midrule
Mesh-Based & \cite{mobilenerf} & 117 FPS & \mygreencheck \\
 \multirow{2}{*}{MLP-Based} & \cite{kilonerf} & 23 FPS &  \multirow{2}{*}{\mygreencheck} \\
 & \cite{kilonerf} w/ Pixel-Reuse~\cite{han2023metavrain} & $>$ 200 FPS &  \\
Low-Rank-Decomp-Based & \cite{reiser2023merf} & 80 FPS & \mygreencheck \\
Hash-Grid-Based & \cite{instant-ngp} & 187 FPS & \mygreencheck\\
3D-Gaussian-Based & \cite{kerbl20233d} & 65 FPS & \mygreencheck \\
\bottomrule
\end{tabular}
}
\vspace{-2em}
\label{tab:absolute_fps}
\end{table}

\textbf{Discussion on Overheads vs. Flexibility.} 
To summarize, the aforementioned (1) reconfigurability and (2) corresponding overheads of the proposed accelerator explain why it (1) outperforms existing accelerators and devices in terms of the overall speedup on varying rendering pipelines and (2) underperforms compared to some dedicated accelerators, e.g.,~\cite{han2023metavrain}, for specific rendering pipelines, respectively. However, as illustrated in Sec.~\ref{sec:exp_hybrid}, with more recent neural rendering pipelines combining multiple representative pipelines for better accuracy vs. efficiency trade-offs, the importance and flexibility of supporting such varying pipelines through the reconfigurability of the proposed accelerator are increasing, compared to merely developing accelerators for a single specific rendering pipeline.
With the aforementioned (1) speedup over commercial devices and (2) flexibility to support varying rendering pipelines compared to prior neural rendering accelerators, our proposed accelerator is the \textbf{first} accelerator to accomplish real-time ($>$ 30 FPS) on-device neural rendering across various representative rendering pipelines, as summarized in Tab.~\ref{tab:absolute_fps}.

\subsection{Evaluation on a \textbf{Hybrid} Rendering Pipeline}
\label{sec:exp_hybrid}
\begin{figure}[t]
    \centering
    \vspace{1.5em}
    \includegraphics[width=\linewidth]{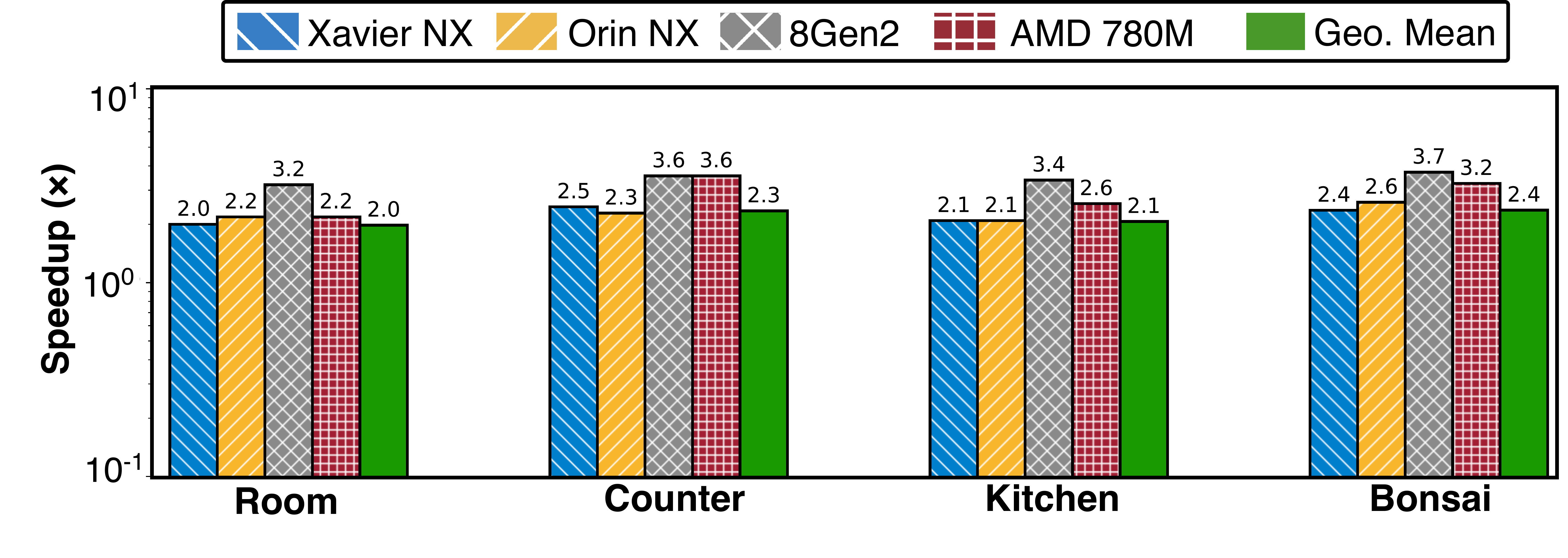}
    \vspace{-2.5em}
    \caption{The speedup achieved by our proposed accelerator over the five baseline devices on a \textbf{hybrid} rendering pipeline~\cite{li2024mixrt} for the four indoor scenes from the Unbounded-360~\cite{barron2022mip} dataset.}
    \vspace{-1.2em}
    \label{fig:exp_hybird}
\end{figure}

Recent advancements in neural rendering have shown a trend toward combining different rendering pipelines to create a hybrid pipeline. This approach merges the merits of each to achieve a better rendering quality-efficiency trade-off. To validate the proposed accelerator's capability to support novel pipelines, we validate its performance on the newly proposed MixRT~\cite{li2024mixrt} neural rendering pipeline, which is a representative hybrid render pipeline that combines mesh-based and hash-grid-based pipelines. As shown in Fig.~\ref{fig:exp_hybird}, we can observe that: (1) the proposed accelerator shows promising universality in the hybrid rendering pipeline, achieving a 2.0$\times$ to 3.7$\times$ speedup across all evaluated baselines; (2) The speedup of our proposed accelerator over baselines is consistent across different scenes, where each scene corresponds to a different model. For example, it consistently achieves a speedup of 2.0$\times$ to 2.6$\times$ compared to the most competitive baselines, Xavier NX~\cite{xnx} and Orin NX~\cite{onx}, across different scenes. The aforementioned evaluation on the hybrid pipeline~\cite{li2024mixrt} illustrates that our proposed accelerator can be adapted to new hybrid rendering pipelines because it does not target a set of specific rendering pipelines but accelerates the identified common micro-operators summarized in Tab.~\ref{tab:common_ops}. 

\begin{table}[!b]\centering
\vspace{-1em}
\caption{A summary of the improvement on the rendering speeds by scaling up the PE array and SRAM sizes. The relative rendering speeds are obtained on the Unbounded-360~\cite{barron2022mip} dataset with the hash-grid-based rendering pipeline~\cite{instant-ngp}.}
\scriptsize
\resizebox{\linewidth}{!}{\begin{tabular}{c|ccc}\toprule
Scale To & 1$\times$ PE Array Size & 2$\times$ PE Array Size & 4$\times$ PE Array Size\\
\midrule
1$\times$ SRAM Size &  \textbf{1$\times$}	&  1.1$\times$ & 1.1$\times$ \\
2$\times$ SRAM Size &  1$\times$ &  \textbf{2$\times$} & 2.2$\times$\\
4$\times$ SRAM Size &  1$\times$ & 2$\times$ & \textbf{4$\times$} \\
\bottomrule
\end{tabular}
}
\label{tab:abl_PE_SRAM_sizes}
\end{table}

\begin{table}[t]\centering
\caption{Comparison of the supported pipelines between our proposed accelerator and other reconfigurable accelerators.}
\scriptsize
\resizebox{\linewidth}{!}{\begin{tabular}{c|ccccc}\toprule
\multirow{2}{*}{Method} & \multirow{2}{*}{Mesh} & \multirow{2}{*}{MLP} & Low-Rank & Hash & 3D\\
&  &  & Decomposed Grid & Grid & Gaussian\\
\midrule
Flexagon~\cite{Flexagon} (NPU)  & \redcross & \mygreencheck & \redcross & \redcross  & \redcross \\
STIFT~\cite{STIFT} (NPU)  & \redcross & \mygreencheck& \redcross& \redcross  & \redcross \\
SIGMA~\cite{qin2020sigma} (NPU)  & \redcross & \mygreencheck& \redcross  & \redcross & \redcross \\
Eyeriss~\cite{eyeriss} (NPU) & \redcross & \mygreencheck & \redcross & \redcross  &\redcross   \\
Plasticine~\cite{prabhakar2017plasticine} (CRGA) & \redcross & \mygreencheck& \mygreencheck & \redcross & \redcross   \\
Ours & \mygreencheck & \mygreencheck &\mygreencheck& \mygreencheck   & \mygreencheck \\

\bottomrule
\end{tabular}
}
\label{tab:comparewcgra}
\end{table}

\subsection{Scaling Analysis on PE Array and SRAM Sizes}
\label{sec:abl_PE_SRAM_sizes}
To explore the optimal hardware configuration of the proposed accelerator's PE array and SRAM sizes, we conduct the scaling analysis to separately scale up the PE array and SRAM sizes, as illustrated in Tab.~\ref{tab:abl_PE_SRAM_sizes}. Specifically, we can observe that when the ratio between the PE array and SRAM sizes is 1:1, the rendering speeds can reach the maximum value given the same areas. This finding can guide us in scaling up the proposed accelerator to handle even larger 3D scenes, as discussed in~\cite{tancik2022block}.

\subsection{Reconfiguration Overhead Analysis}
\label{sec:reconfiguration_overhead}

In the analysis of the reconfigurable overhead of our proposed accelerator, two aspects are considered: \textbf{(1) Efficiency Impact}: Implementing certain algorithms may reduce the efficiency of others. For instance, to execute GEMM operations within our proposed accelerator, data must pass through a buffer before reaching ALUs, potentially lowering throughput since this is an additional stage compared to the vanilla systolic array\cite{tpu}. As a result, for the MLP-based rendering pipelines that consist of massive GEMM operations, dedicated accelerator MetaVRain~\cite{han2023metavrain} is \textcolor{black}{2.8}$\times$ more energy efficient than our proposed accelerator, when benchmarking the energy needed for rendering each pixel. However, the proposed accelerator maintains real-time rendering speed in all pipelines despite some performance trade-offs, as illustrated in Tab.~\ref{tab:absolute_fps}. \textbf{(2) Module Utilization}: Not all modules in the proposed accelerator are used for each micro-operator. For instance, executing GEMM leaves the special function units idle. Thus, we leverage power and clock gating help conserve energy and minimize the impacts of unused modules.

%% file: sections/6-RelatedWorks.tex
\section{Related Works}

\subsection{Neural Rendering Accelerators}
Existing accelerators for neural rendering acceleration~\cite{han2023metavrain,li2022rt,li2023instant,fu2023gen,Lee2023Neurex,mubarik2023hardware,yuan20240} employ three major strategies: (1) \textbf{Memory Access Optimization}:~\cite{mubarik2023hardware, li2023instant, li2022rt} address irregular memory access, a significant efficiency bottleneck in hash-grid-based and low-rank-decomposed-grid-based neural rendering pipelines; (2) \textbf{Space Tiling}: To mitigate memory bank conflicts,~\cite{li2022rt, Lee2023Neurex} implement space tiling to encourage data access within a smaller, more manageable memory space; (3) \textbf{Algorithm-Level Optimizations}:~\cite{fu2023gen,han2023metavrain} focus on customizing algorithms for specific neural rendering pipelines to make them more hardware-friendly. Despite these efforts, all the works mentioned focus on optimizing single neural rendering pipelines. Currently, no existing accelerator can support multiple or all typical rendering pipelines, unlike the approach we propose in this work.
We note two recent works targeting specific neural rendering pipelines: GSCore~\cite{lee2024gscore} for 3D Gaussian and CICERO~\cite{feng2024cicero} for Hash Grid. In comparison with GSCore in the 3D Gaussian-based pipeline, our approach is only 20\% slower, i.e., GSCore achieves a 15× speedup over XNX~\cite{xnx}, while we achieve a 12× speedup. When comparing with CICERO in the hash-grid-based pipeline, our approach is 14\% slower, when scaling to the same number of MAC units.

\subsection{CGRAs, FPGAs, and NPUs}
Coarse-Grained Reconfigurable Architecture (CGRA) and Field Programmable Gate Arrays (FPGA) are well-known for their flexibility in supporting various workloads~\cite{li2022coarse,taras2019impact,tan2020opencgra,emani2021accelerating,sankaralingam2022mozart,voitsechov2014single,nowatzki2016pushing,prabhakar2017plasticine,gobieski2021snafu,qiu2022tram,yuan20240}. Compared to prior CGRA/FPGA works, our proposed accelerator distinguishes itself as a dedicated solution for neural rendering applications. Specifically, our proposed accelerator achieves a 25$\times$ speedup over \cite{qiu2022tram} on MLP-based pipelines because \cite{qiu2022tram} focuses on vector computations with bulk memory access for vectors or matrices, whereas our proposed accelerator accommodates the massive varying scalar computations in neural rendering. The scalar computations result in more irregular memory access due to complex positional encoding and unpredictable input camera poses, as analyzed in \cite{li2023instant}. Meanwhile, compared to \cite{yuan20240}, an FPGA-based neural rendering accelerator, our proposed accelerator achieves a 15$\times$ speedup and an \textcolor{black}{10}$\times$ improvement in energy efficiency on hash-grid-based pipelines, thanks to its more flexible input/reduction data networks (as detailed in Sec.~\ref{sec:arch:reconfigurable_data_networks}) designed for the aforementioned irregular memory access. Neural Processing Units (NPUs) are designed to primarily support traditional neural network operators, but they do not support graphics operators~\cite{qin2020sigma, eyeriss}. Tab.~\ref{tab:comparewcgra} compares the supported pipelines of this work with those of other traditional accelerators.

%% file: sections/7-Conclusion.tex
\section{Conclusion}
We present a unified neural rendering accelerator designed to overcome key challenges in real-time, on-device neural rendering. Our research alleviates the gap between the desired real-time rendering speeds for immersive interactions across diverse applications and the slow rendering speeds achieved by existing neural rendering hardware. Specifically, we propose a unified accelerator with a reconfigurable hardware architecture that can dynamically adapt to various pipeline requirements, demonstrating significant speedups over existing accelerators while maintaining power efficiency.
To the best of our knowledge, our proposed accelerator stands out as the first framework to realize real-time neural rendering on-device for a range of rendering pipelines, while also meeting the power efficiency standards of edge devices.

%% file: sections/8-Acknowledgment.tex
\section*{ACKNOWLEDGMENTS}
This work was supported by the National Science Foundation (NSF) Computing and Communication Foundations (CCF) program (Award ID: 2434166 and 2312758), the Department of Health and Human Services Advanced Research Projects Agency for Health (ARPA-H) under Award Number AY1AX000003, and CoCoSys, one of the seven centers in JUMP 2.0, a Semiconductor Research Corporation (SRC) program sponsored by DARPA. The measurements on Snapdragon® Mobile Processors 8 Gen 2 were conducted using the Qualcomm Innovators Development Kit (QIDK), supported by Qualcomm.